\documentclass{elsart}
\input epsf
\usepackage{graphicx}

\def\gr{$\gamma$-ray \,}

\begin{document}

~~~~~~~~~~~~~~~~~~~~~~~~~~~~
~~~{\em Astroparticle Physics, accepted for publication} 

\runauthor{}
\begin{frontmatter}
\title{5@5 - a 5 GeV energy threshold 
array of imaging atmospheric Cherenkov  
telescopes at 5 km altitude}

\author{F.A. Aharonian, A.K. Konopelko, H.J. V\"olk}
\address{MPI f\"ur Kernphysik, D-69029 Heidelberg, Germany}
\author{H. Quintana}
\address{Astronomy and Astrophysics Department, 
Pontificia Universidad Catolica de Chile,
Casilla 306, Santiago 22, Chile}
\begin{abstract}
We discuss the concept and the performance of a 
powerful future ground-based astronomical instrument,   
5@5  - a {\bf 5} GeV energy threshold stereoscopic  array of several large 
imaging atmospheric Cherenkov telescopes installed {\bf at} a very 
high mountain elevation of about {\bf 5} km a.s.l. -  
for the study of the  $\gamma$-ray sky at energies  from approximately 
5 GeV  to 100 GeV, where the capabilities of  both the current space-based 
and ground-based  $\gamma$-ray  projects are quite limited.
With its potential to detect the ``standard''  EGRET 
$\gamma$-ray sources with spectra extending beyond several GeV
in exposure times from 1 to  $10^3$ seconds, 
such a detector may serve as an ideal 
"Gamma-Ray Timing Explorer" for  the study of transient 
non-thermal phenomena like $\gamma$-radiation  
from  AGN jets, synchrotron flares of microquasars, 
the high energy (GeV) counterparts of  Gamma Ray Bursts, {\it etc.} 
5@5  also would allow  detailed $\gamma$-ray spectroscopy 
of persistent nonthermal  sources like pulsars,  
supernova remnants, plerions,  radiogalaxies, and others,  with
unprecedented  for $\gamma$-ray astronomy photon statistics.  
The existing technological achievements in the design and 
construction of multi (1000) pixel, high resolution imagers, 
as well as of large, 20 m diameter class multi-mirror dishes with  
rather modest optical requirements, would allow the construction  
of such a detector in the foreseeable future,  although in the 
longer terms from the point of view of ongoing projects of 100 GeV 
threshold IACT arrays like H.E.S.S. which is in the build-up phase. 
An ideal site for  such an instrument could be a high-altitude, 
5 km a.s.l or more, flat area with a linear scale  of about 100 m  
in a very arid mountain  region in the Atacama desert of  Northern Chile.
\end{abstract}

\end{frontmatter}
\section{Introduction}

The high detection rate, the ability of effective separation of 
electromagnetic and hadronic showers, and the  good accuracy of 
reconstruction of the direction of primary $\gamma$-rays are three
remarkable features of the imaging atmospheric Cherenkov telescope
(IACT) technique (see e.g. \cite{Cawley,Hillas,Fegan}). 
The recent detections  of TeV $\gamma$-rays from 
several galactic and extragalactic objects (see e.g. \cite{Weekes99})
provide the basis  for the further 
development of {\em ground-based}  gamma-ray astronomy.
 
Qualitative improvements  of the IACT technique in the next few
years will most probably  be linked to {\em stereoscopic} observations 
of $\geq 100 \, \rm GeV$ $\gamma$-rays  \cite{AhAk}. 
This approach not only allows an unambiguous
determination of the energy and of the arrival direction 
of $\gamma$-ray primaries 
on an {\it event-by-event} basis,  but also significantly  improves the 
efficiency of rejection of  hadronic showers produced 
by cosmic rays \cite{array}, as it was  recently 
demonstrated by the HEGRA system  of 5 imaging telescopes  
operating in  the energy region from 
500 GeV to 20 TeV \cite{Daum,hegra_MC}. 
 
One of the important issues for future detectors is the choice of the
energy region based on two principal arguments: (a) astrophysical 
significance (goals) and (b)  the experimental 
feasibility/reliability (cost). If one limits the 
energy region from above to a 
relatively modest energy threshold around 100 GeV, 
then the  performance 
of IACT arrays and their practical implementation can be predicted with 
confidence. In practice, an energy threshold of $\sim 100$ GeV 
can be achieved  by a stereoscopic system of 
IACTs consisting of 10~m diameter 
class optical mirrors and equipped with 
high resolution cameras, based on
conventional photo-multipliers (PMTs). 
Currently three such arrays are under 
development/construction  in Australia (CANGAROO-3 \cite{CANG}),
in Namibia (H.E.S.S. \cite{HESS}), and 
in Arizona (VERITAS \cite{VERITAS}). 
With their superior angular  resolution of 
several arcminutes and  energy flux sensitivity close 
to $10^{-13} \, \rm erg/cm^2 s$, these projects  
perfectly suit to the energy range from 50 GeV to 10 TeV, 
which,  from the point of view of 
scientific motivations and the potential 
astronomical targets, can be considered as 
a spectral domain  in its own right.    

On the other hand, it is expected that  the next generation major 
satellite $\gamma$-ray mission  GLAST 
(see e.g. \cite{Bloom,GehMich}),
the successor of the EGRET instrument aboard the  
Compton Gamma Ray Observatory,  will extend the exploration of 
the $\gamma$-ray sky up to 100 GeV. Thus the gap between  
space-based and ground-based $\gamma$-ray instruments will eventually 
disappear. It should be noticed, however, that  
in many cases this statement has a rather conditional or even 
symbolic character. Although at GeV
energies GLAST will improve the EGRET sensitivity by almost
two orders of magnitude, the capability of GLAST (and likely that 
of any post-GLAST space-based project) at energies well beyond 
10 GeV will be quite limited because of the limited 
detection area.  
 
This circumstance justifies recent activities to reduce the energy 
threshold of atmospheric Cherenkov detectors below 100 GeV. 
Currently two low threshold projects, CELESTE in 
France \cite{celest} and STACEE in USA \cite{stace}, based on 
the concept of conversion of existing  solar power plants into 
atmospheric Cherenkov telescopes, 
are in their final stage of realization.  Although 
this technique allows  in principle a  reduction of the energy 
threshold down to  20-30 GeV (see e.g. Ref. \cite{Pare}),
the challenge remains to reach an adequate  
detection efficiency of $\gamma$-rays 
at such low energies.
Another approach has been  proposed  by the  MAGIC collaboration
with a single  imaging Cherenkov  telescope  
having  a large, 17 m  diameter reflector. Equipped with a standard 
PMT-based high resolution camera this telescope 
is expected to allow effective 
detection of $\gamma$-rays at energies above 30 GeV \cite{magic}.

With some exceptions, the GeV  \gr sources ($E \geq 0.1\, \rm GeV$) 
are expected to  be quite different from  TeV sources 
($E \geq 0.1 \, \rm TeV$). The proximity of the intermediate domain 
below 100 GeV  to the energy range covered by EGRET suggests 
that many objects established  as GeV emitters have a good chance to be 
detected  also by the above mentioned ground-based instruments. 
This argument, however, cannot yet guarantee definite success.
Indeed,  although the two largest $\gamma$-ray source populations 
identified by EGRET, radiopulsars  and distant AGN, do not show 
a significant steepening or cutoff up to 10 GeV, the theoretical studies 
of \gr production and absorption conditions in these objects,
as well as  rather general phenomenological considerations  
predict cutoffs in the energy spectra around 10 GeV or less. 
In addition, for any reasonable model of the diffuse 
extragalactic cosmic background radiation, 
we should expect sharp cutoffs in the spectra of  
distant extragalactic objects 
with redshift $z \sim 1$ at energies as low as 
50 GeV (see e.g. \cite{primack}). This implies that for 
the study of cosmologically distant sources, like the GeV   
blazars discovered  by EGRET,  or Gamma Ray Bursts (GRBs),  the energy 
threshold of  the detectors should be less  than 10 GeV at which energy
the Universe is most likely transparent up to at least $z \simeq 3$.
An instrument like GLAST, operating effectively in the  
0.1 to 10 GeV energy region,  nicely suits this task.   
In particular, it is expected that the number of 
AGN  that GLAST will detect could exceed 
several thousands \cite{GehMich}. 
At the same time, the  relatively small detection area of GLAST, 
$A_{\rm eff} \simeq 0.8 \, \rm m^2$,  limits the potential 
of this instrument for detailed studies of the 
temporal and spectral 
characteristics of highly variable  $\gamma$-ray sources
like blazars, which have  variability timescales less 
than a few hours, or of solitary events like GRBs with a duration 
of $10^{-2}$ to $10^3$ seconds. In this regard,
GLAST can hardly match the performance of current X-ray detectors 
that have similar detection areas but operate 
in a regime of photon fluxes that exceed  the fluxes of 
MeV/GeV $\gamma$-rays  by many orders of magnitude. 

The idea of a  
``Gamma-ray  timing explorer''
to study transient $\gamma$-ray phenomena  with an adequate
photon detection rate  motivated, to a large extent,
the present investigation. It concerns the possible  extension of 
the domain of ground-based Cherenkov technique 
with its huge  detection area of   
$10^4$ to $10^5 \, \rm m^2$  
down to energies of several GeV. 
We shall argue  that  such a  goal could be best achieved 
by stereoscopic systems of several large, 20 m class
imaging atmospheric Cherenkov telescopes located at very high, 
$H \sim 5 \, \rm km$, mountain altitude.

\section{Concept of an IACT array with  5 GeV threshold}

The concept of stereo imaging is based on the simultaneous 
detection of a single air shower in different projections by at least two 
telescopes, separated at a distance comparable with
the ``effective radius''   
$R_{\rm C} \sim 100 \, \rm m$ of the Cherenkov light pool. 
The stereoscopic approach allows {\em (i)} unambiguous and 
precise reconstruction of shower parameters on an 
{\it event-by-event} basis, {\em (ii)} superior rejection of 
hadronic showers, and {\em (iii)} effective suppression of the 
background light from different sources - the {\em nigh sky background} 
(N.S.B.), local muons, etc. \cite{array}. All these three 
advantages over single IACTs have been 
convincingly demonstrated at TeV energies by  the HEGRA 
IACT system \cite{Daum,hegra_MC}.

Compared with single (``stand alone'') telescopes,
which can adequately measure the shower inclination 
in the direction perpendicular to the plane containing 
the telescope axis,  but poorly in the in-plane direction,
the stereoscopic approach allows full reconstruction 
of the arrival direction of individual $\gamma$-ray showers.
Apart from the good directional information,
stereoscopic IACT systems  make use of the fact that
the Cherenkov images of a shower detected by several different, 
spatially separated telescopes,  are only {\it partially} 
correlated.  Therefore the stereoscopic measurements 
significantly improve the efficiency of rejection of 
hadronic (background) showers at both the hardware (trigger) 
and the software levels. The only disadvantage of the 
stereoscopic approach is a non-negligible loss in the
detection rate because of the overlap of the shower detection
areas of individual telescopes located from each other at distances
$\leq 2R_{\rm C}$. However, this loss of statistics is compensated,
especially for steep spectra of primary $\gamma$-rays, by a 
significant reduction of the energy threshold of the telescopes 
operating in coincidence mode.            

At $\gamma$-ray energies above 100 GeV stereoscopic
IACT arrays do provide an excellent angular resolution 
of about $0.1^{\circ}$ or less, and a ``gamma/hadron'' 
separation efficiency (including the hadron rejection at the 
trigger level) of 1000:1. 
This improves the flux sensitivity dramatically
compared with the sensitivity of single telescopes.   
The efficiency of the imaging technique 
is somewhat lower at energies  below 100 GeV. In particular, 
the Cherenkov images become less elongated and less regular. 
In practice this introduces significant 
uncertainties in the reconstruction of image parameters. 
Even so, below it will be shown that the performance of the 
stereoscopic imaging remains adequately high even below 
10 GeV.

The {\em effective energy threshold} of IACTs
is basically determined by two conditions:
{\em (i)} the number of photoelectrons 
in the image should be sufficient
for an appropriate image analysis;
typically, ``good imaging'' requires 
$n_{\rm ph.e.}^{\min} \sim 50-100$ electrons\footnote{For example, 
in the case of the HEGRA stereoscopic IACT system 
the minimum number of photoelectrons  
corresponding to the  showers classified as ``high quality events'', i.e. 
the showers which are accepted for the further 
image analysis, is close to 40 electrons 
per telescope \cite{hegra_MC}.},   
{\em (ii)} the accidental trigger rate introduced by the 
N.S.B. should not exceed the detection rate of $\gamma$-rays.
For  an ideal IACT the first  condition in principle 
should dominate over the second, technical condition.

The number of photoelectrons detected by the imager - 
a multi-pixel camera placed in the focal plane 
of the mirror -  depends on (i) the telescope photo-electron 
response (or aperture), 
$S_{\rm ph.e.}=S_{\rm mir} \cdot \xi_{\rm ph \rightarrow e}$ 
(where $S_{\rm mir}$ is the geometrical
area of the mirror, and $\xi_{\rm ph \rightarrow e}$ is the  
photon-to-photoelectron conversion factor), and (ii) the density 
of optical Cherenkov photons $\rho(R,E)$ produced at the 
typical  distance $R \sim 100 \, \rm m$ from the axis of the shower 
initiated by a primary  $\gamma$-ray of energy $E$.
  
The density of the Cherenkov light at an elevation of $H=5$ km a.s.l.
within 100 m of the shower core  produced by a primary  
$\gamma$-ray photon of energy  $E=5 \, \rm GeV$ is close to  
$1 \, \rm photon/m^2$. Correspondingly, the number of photoelectrons 
detected by a telescope  with aperture $S_{\rm ph.e.}$ is equal to 
$n_{\rm min} \approx 1 \ S_{\rm ph.e.}$. Thus, for reduction of  
the energy threshold down to 5 GeV,  the telescope aperture 
$S_{\rm ph.e.}$ should be as large as $50 \, \rm m^2$, assuming
that the minimum number  of electrons required for 
an image analysis $n_{\rm ph.e.}^{\rm min} \sim 50$. 
Detailed Monte-Carlo calculations presented below
confirm that the energy threshold of such an instrument,
determined as  the energy at which the differential 
$\gamma$-ray detection rate reaches to its maximum,
could indeed be as low as 5 GeV. For conventional aluminized 
optical mirrors and PMT-based cameras with
typical conversion factor  $\xi_{\rm ph \rightarrow e} \sim 0.15-0.2$,
$S_{\rm ph.e}=50 \, \rm m^2$  would require 
a large optical reflector of approximately 20 m diameter. 
With the successful development of novel, fast (nsec)
detectors of optical radiation with a quantum efficiency exceeding 
$50 \,  \%$,  the energy threshold of the telescopes could be pushed 
further down to 2 or 3 GeV, which is an absolute limit determined by the 
minimum energy of secondary electrons capable of producing Cherenkov
light in the upper atmosphere.

The Cherenkov light density increases monotonically with 
elevation. Thus, for the given telescope configuration,
the installation of an IACT array at very high altitude
would allow a straightforward and  unconditional reduction 
of the energy threshold.
At the same time  the increase of the telescope aperture leads 
not only to a proportional increase of  number of
photoelectrons registered  per shower, but also to an 
increase of the accidental rate caused by the N.S.B. 
Thus, a significant gain from large aperture telescopes, 
with the  ultimate goal to operate the telescopes in 
the linear regime,  $E_{\rm th} \sim 1/S_{\rm ph.e}$, 
can be achieved only through effective suppression 
of the N.S.B. Otherwise the reduction  of the energy threshold  
would be rather slow, $E_{\rm th} \sim 1/S_{\rm ph.e}^{1/2}$, 
and therefore would be difficult to  justify economically.  

In the imaging technique, the accidental events introduced by 
the N.S.B are  suppressed by a trigger condition
that requires signals above a threshold $q_0$ (photoelectrons) 
in $m \geq 2$ adjacent pixels.
A smaller pixel size not only reduces significantly 
the noise level due to the  N.S.B., but it 
allows higher trigger multiplicity as well. 
At the same time  the optimal size of the pixel is determined by the 
condition of ``good imaging'', which implies 
that the area  which is  covered by the minimum number of 
pixels (typically 10) used in the image analysis,  should 
not exceed the characteristic size of the image. At 
$E \sim 10 \, \rm GeV$, the image area  of electromagnetic 
showers is less than $0.2$ square degree. Therefore at such low energies          
high resolution cameras with a pixel size  
close to $0.1^{\circ}$ provide  an adequate
imaging quality.  Although such small pixels 
allow also significant reduction of the N.S.B.,   
it appears that for a single IACT with aperture $50 \, \rm m^2$, 
even a  pixel size of $0.1^{\circ}$ is still
not sufficient to operate the telescope at the minimum possible 
energy threshold, i.e. in the regime
when the detection threshold is determined by the
Cherenkov light amplitudes rather than by 
the  N.S.B. noise. Further reduction of the pixel size 
makes the design of the imager with many thousands of channels 
technically very difficult. A solution of this dilemma 
is the  stereoscopic mode of observations
which offers a more feasible and economic approach.
It requires simultaneous detection of a shower by 
at least two   telescopes.
Because of the flat lateral distribution of 
the Cherenkov radiation from electromagnetic  showers, this 
requirement does not effect the $\gamma$-ray detection 
efficiency (if the distance between telescopes does not 
significantly exceed 100 m), but significantly reduces the accidental rate
caused by the N.S.B. Below we will indeed show that the suppression 
of the N.S.B. by a stereoscopic system of IACTs with camera
pixel size $\sim 0.1^\circ$ is  sufficient to
operate the $50 \, \rm m^2$  aperture telescopes
in the energy regime below 10 GeV.
 
The arrangement of an IACT array, in particular 
the number of telescopes, and the spacing between them, can 
be understood from the following simple considerations.
The Cherenkov light pool on the ground produced by primary $\gamma$-rays
of energies $E \geq 100 \, \rm GeV$ has a flat radial 
distribution with a pool radius of approximately 100 m. Hence the optimal  
spacing between the telescopes should be of the order 
of 100 m. A significantly smaller spacing reduces not 
only the detection  area, but also the quality of images.
A spacing of significantly more than 100 m reduces 
the coincidence rate dramatically, especially at low energies, and thus 
increases the energy threshold.  
The  special investigation carried out for 100 GeV class 
telescopes \cite{array,optim}, generally 
confirms this simple conclusion. We expect that this 
should be the case also for the sub-10 GeV array, although 
detailed Monte-Carlo simulations are needed for 
optimization of the arrangement of reflectors in this array.
Here we consider a  reasonable arrangement 
with a spacing of 100 m. The question of optimization 
of the spacing will be discussed in a separate paper.     
The minimum number of telescopes is determined from the
consideration that at least 3 stereoscopic views are needed
to reconstruct the shower parameters  reliably \cite{array}.
For clarity, we  assume here a square baseline with  telescopes 
at the four corners, placed  at an altitude $H$ of 5 km a.s.l.
In order to have more homogeneous 
coverage of distances of impact parameters,
especially for showers with cores outside the square,
we assume in addition a further (5th) telescope in the center of the square.
The main objective of this paper is
to study the basic performance of a 5 GeV IACT array 
placed  at very high altitude. The optimization of the 
arrangement of the array  is outside 
of the scope of this paper, but we believe that the suggested 
layout is not far from the optimum design.

\section{Basic characteristics of Cherenkov radiation 
induced by sub-10 GeV gamma-rays}

The results discussed in this section have been obtained 
by using the ALTAI code \cite{altai} that simulates  
electromagnetic and hadronic showers and their Cherenkov 
radiation in the Earth's atmosphere.
This code has been used before for calculations of the performance 
of  the HEGRA system of IACTs \cite{hegra_MC}, 
as well as for studies of the expected characteristics  
of the new  generation ``100 GeV - threshold''  IACT arrays 
\cite{array} that are the basis of the H.E.S.S. (High Energy Stereoscopic System)
project in Namibia \cite{HESS}.  The predictions for  
the HEGRA IACT system  have been thoroughly checked 
using the detected  hadronic (background) showers produced 
by cosmic rays  with a relatively well known energy spectrum and mass 
composition \cite{hegra_MC,proton_paper}. More importantly, the calculations 
of the characteristics of the instrument for primary $\gamma$-rays,
in particular the  lateral distribution of the Cherenkov light,
which determines the $\gamma$-ray detection area,    
the angular resolution, and the  gamma/hadron  separation
efficiency, have been experimentally confirmed by 
the HEGRA collaboration using 
$\gamma$-ray  data obtained during the active 
state of Mkn 501 in 1997 \cite{hegra_lat,mkn501}.
This  extraordinary high state of the source,  
with a duration of several months,  resulted in approximately 40,000  
$\gamma$-rays  in the energy interval from 500 GeV to 20 TeV,    
detected by the HEGRA IACT system under  almost background-free 
conditions \cite{mkn501}. 

In this paper we are interested in the possibility of detecting   
primary $\gamma$-rays over an interval of very low energies
from several GeV to 100 GeV,  by a  system of IACTs located  at 
$H \sim 5 \, \rm km$ a.s.l. 
Both conditions are rather extreme, and exhibit 
features which differ significantly  from traditional studies 
of air showers. In particular,  at energies of primary 
$\gamma$-rays  below 10 GeV,  we deal with Cherenkov radiation 
from only a  handful of first generation electrons, while at TeV energies 
the Cherenkov radiation is contributed  by a large number  of 
electrons produced during  the full cascade development.
At such low energies we  have  therefore to expect more fluctuations of the   
parameters characterizing the showers and their Cherenkov 
radiation. In particular, the Cherenkov images of sub-10 GeV 
showers are expected to have a less regular shape, 
as well as to be more strongly affected 
by the geomagnetic field,  as compared with the showers 
in the  $\geq 100 \, \rm GeV$ energy region.
Also, both the very low energy domain and the very high elevation
of the location of the suggested IACT array imply not only better transparency  
conditions for the Cherenkov radiation, but also 
a non-negligible  reduction of the light produced above the telescopes 
from  hadronic showers.
All these effects add new features to the  characteristics of the 
Cherenkov radiation of air showers.  

\subsection{Lateral and longitudinal development of showers}

The lateral distribution of Cherenkov radiation from  air showers,   
induced by $\gamma$-rays of energy 10 GeV and by protons of energy
100 GeV,  are shown in Fig.~1 at 3 different observational  
levels - $H=$2.2 km, 3.5 km and 5 km.
It is seen that within the radius of 100 m from the shower core 
which typically determines the detection area of IACTs, 
the density of Cherenkov radiation from  $\gamma$-ray induced showers 
has {\em (i)} a rather flat lateral distribution,  
and  {\em (ii)} increases significantly with 
elevation above sea level. In particular, the rise from 2.2 km to 5 km 
results in an increase of  the density of the Cherenkov light by a factor
of 2 to 3 (Fig.~1a).  This  elevation effect  is less 
pronounced for  proton-induced showers 
($\leq 50$ per cent at $r=100 \, \rm m$; see Fig.~1b).
This implies that the choice of very high elevation for the operation of
telescopes would allow a  significant reduction of the energy threshold,
by a factor of $\sim 2.5$,  as well as 
a noticeable improvement of the background rejection 
of hadronic showers.

\begin{figure}[htbp]
\begin{center}
\includegraphics[width=0.45\linewidth]{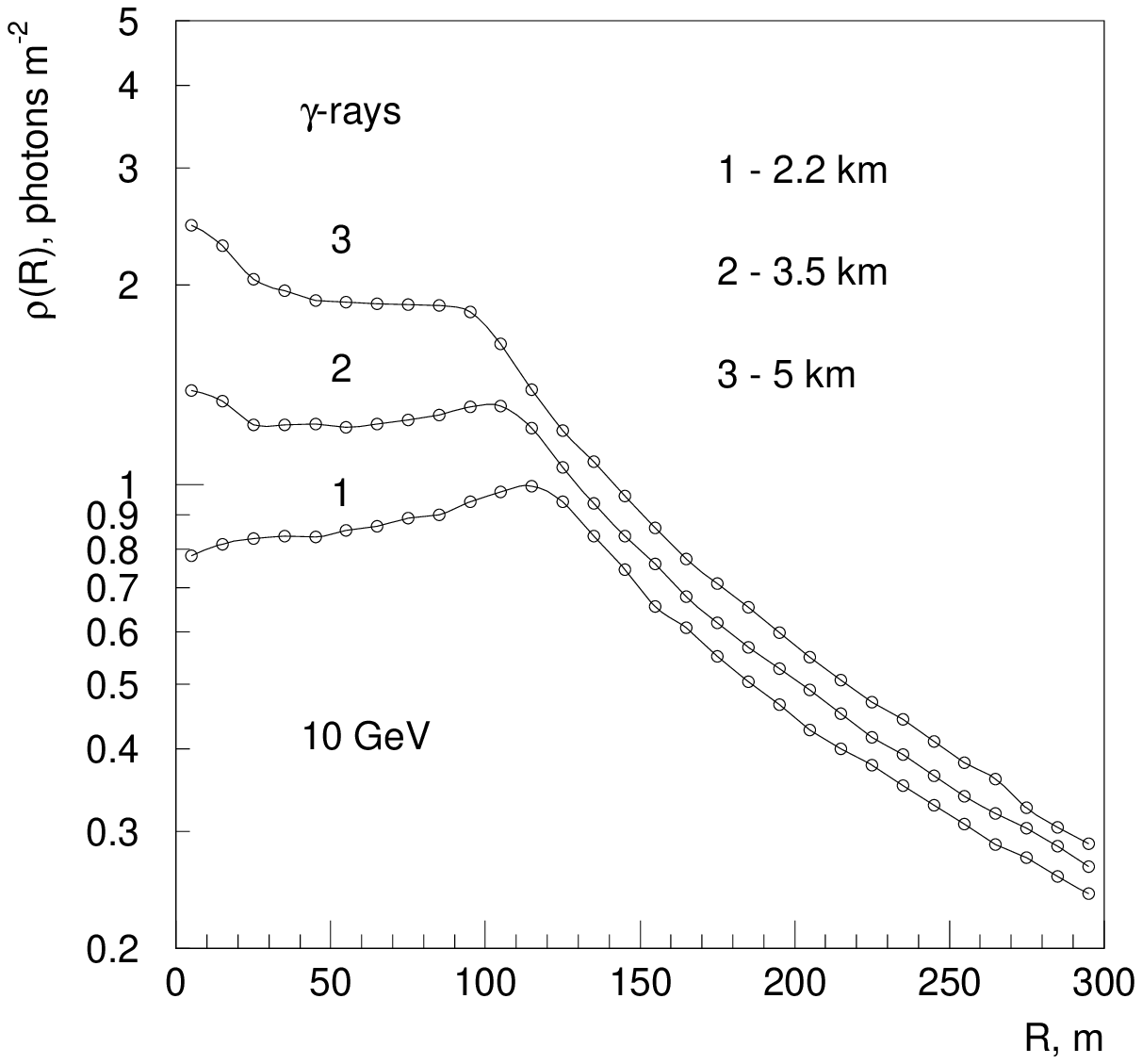}\hspace{5mm}
\includegraphics[width=0.45\linewidth]{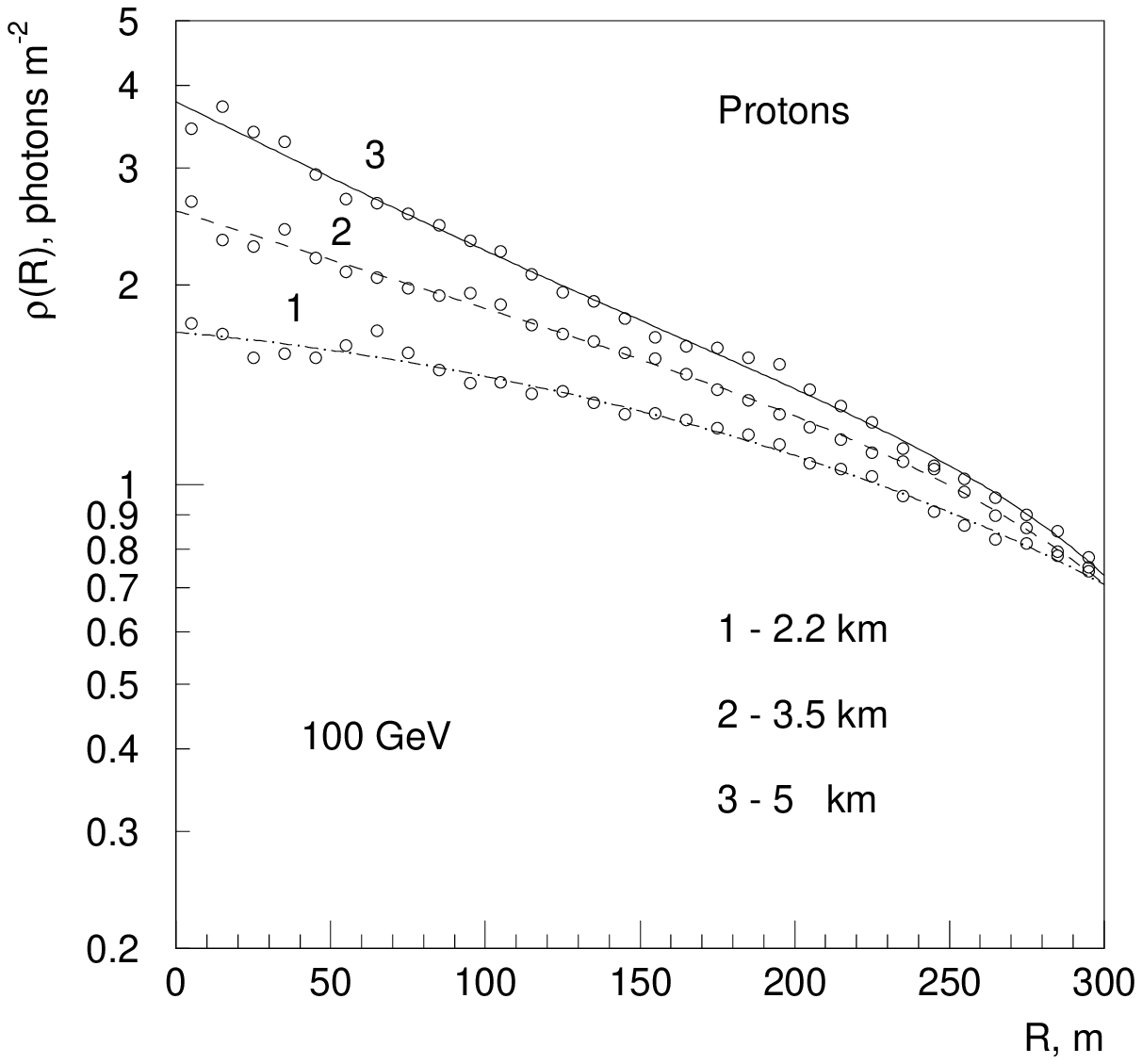}
\caption{Lateral distribution of the Cherenkov light density from 
{\bf (a)} a primary $\gamma$-ray photon of energy 10 GeV
(left) and {\bf (b)} a cosmic ray proton of energy 100 GeV (right)
at three different observation levels above sea level: 
2.2 km, 3 km, and 5 km.}
\end{center} 
\end{figure}

The slow increase  of the Cherenkov light density from  hadronic showers
with the rise of elevation is explained by the effect of deep penetration
of the ``$\pi^{\pm}$-branches'' (sub-showers) of the cascade into the atmosphere 
(Fig.~2), which results in the production of a non-negligible amount of  
Cherenkov light below the observation  level at $H \sim 5 \, \rm km$.
This effect is quantitatively demonstrated 
in Fig.~3 where the longitudinal  distributions of the Cherenkov 
light, produced by a 10 GeV $\gamma$-ray photon and a 100 GeV proton,
are  shown.   
%
\begin{figure}[htbp]
\begin{center}
\includegraphics[width=0.75\linewidth]{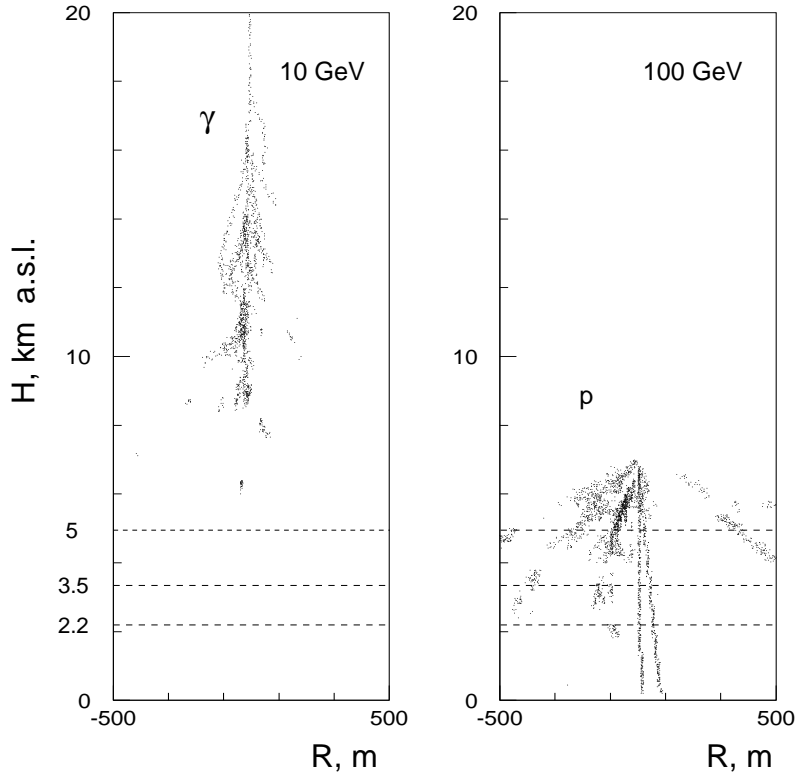}
\caption{Longitudinal development of $\gamma$-ray and proton 
showers as seen in Cherenkov light radiated by the secondary electrons.
The more irregular proton showers lead to less slender images than 
$\gamma$-ray showers.}
\end{center} 
\end{figure}

The low energy threshold  itself  already 
leads, in fact  more strongly  than the  elevation effect,  
to a significant improvement of the gamma/hadron  ratio. 
This effect, caused by the  reduction of the yield of Cherenkov 
light by sub-100 GeV hadronic showers (see e.g. \cite{Pare}) 
is demonstrated  in Fig.~4 for $H=$5 km a.s.l. The density of 
Cherenkov light in the electromagnetic showers is approximately 
proportional  to the $\gamma$-ray energy  down to $E \sim 10 \, \rm GeV$,
the dependence becoming  a bit stronger 
below 10 GeV  (Fig.~4a).  At the same time,  the density of 
\begin{figure}[htbp]
\begin{center}
\includegraphics[width=0.6\linewidth]{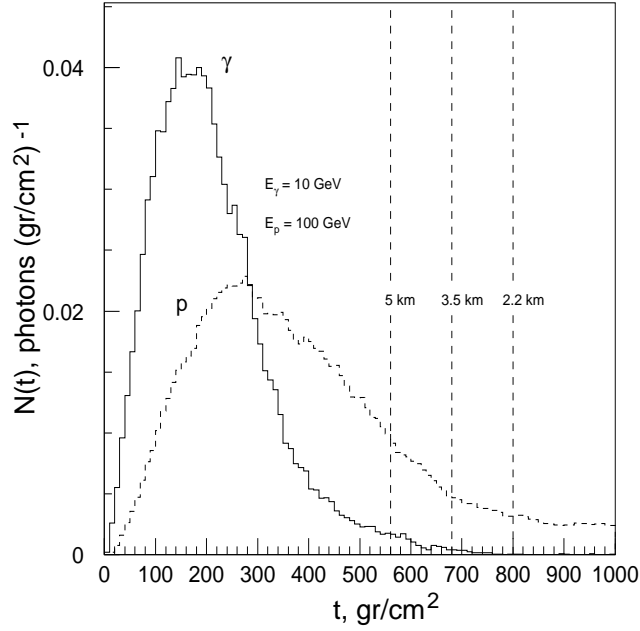}
\caption{Longitudinal distribution  of the Cherenkov light 
produced by $\gamma$-ray and proton-induced showers.}
\end{center} 
\end{figure}
Cherenkov light from  hadronic showers drops significantly 
faster with reduction of the primary energy (Fig.~4b).
In Fig.~5 we show the ratio of densities of the 
Cherenkov light $\rho^{(\gamma)}_{100}(E_\gamma)$ and
$\rho^{(\rm p)}_{100}(E_{\rm p})$
produced by $\gamma$-rays and protons at 
100 m distance from the shower core.
\begin{figure}[htbp]
\begin{center}
\includegraphics[width=0.45\linewidth]{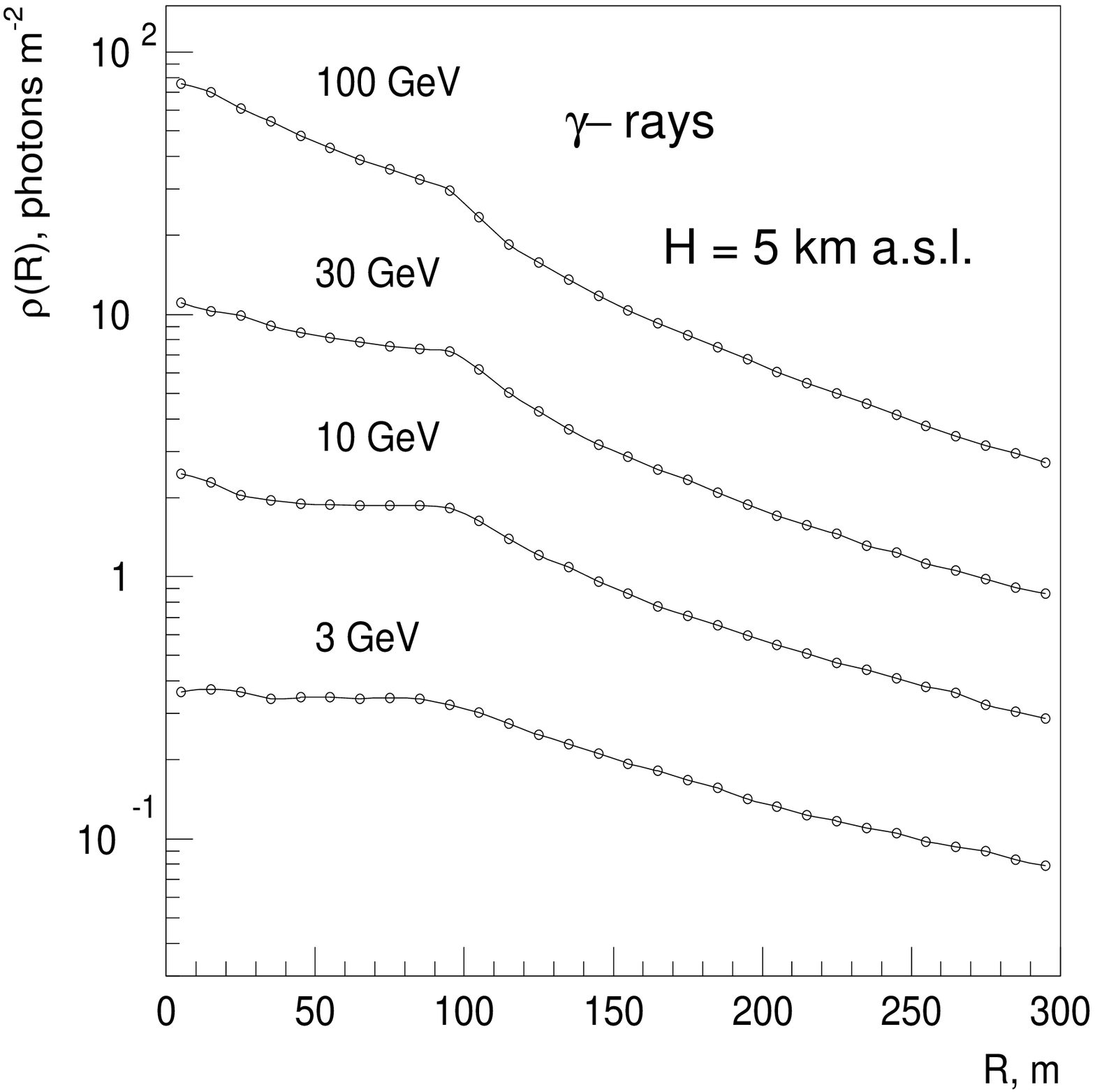}\hspace{5mm}
\includegraphics[width=0.45\linewidth]{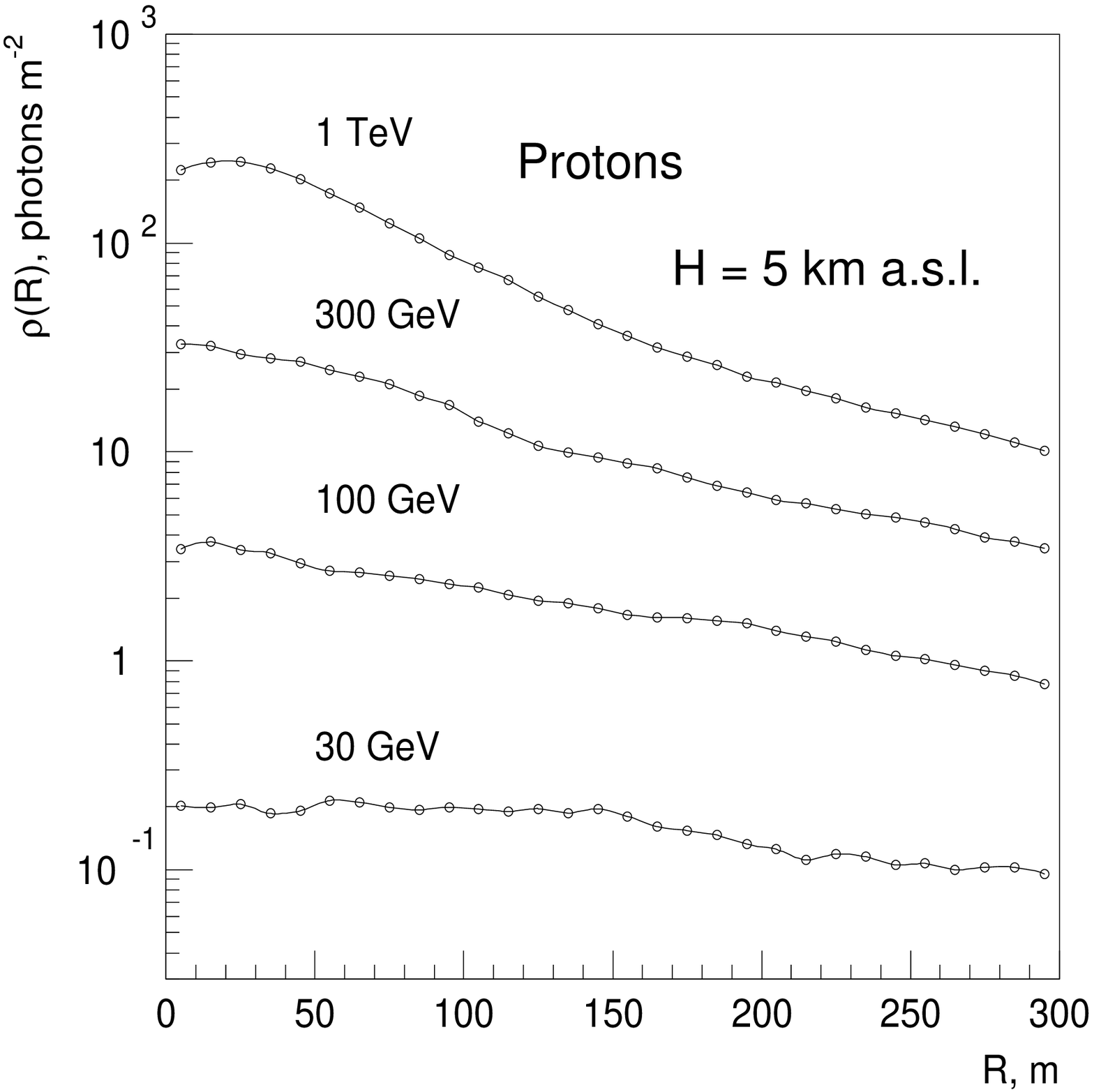}
\caption{Lateral distribution of the Cherenkov light density in
{\bf (a)} $\gamma$-ray (left) and 
{\bf (b)} proton (right) induced  showers at an
altitude of 5 km a.s.l. The energies of 
primary particles are shown at the curves.}
\end{center} 
\end{figure}
Dramatic rise of this ratio
below 100 GeV  implies  
a monotonic increase of the gap between the 
energies of electromagnetic and hadronic showers 
that produce the same amount of Cherenkov light.
This effect is demonstrated in Fig.~6. The solid line corresponds
to the relation between energies of $\gamma$-rays 
$E_\gamma$ and protons $E_{\rm p}$
defined from  equation 
$\rho^{(\gamma)}_{100}(E_\gamma)=\rho^{(\rm p)}_{100}(E_{\rm p})$.
For comparison, the line ``$E_\gamma=E_{\rm p}$''
is also shown.  It is seen, in particular, that the
gap between $E_\gamma$ and $E_{\rm p}$,
which  is close to $\approx 2$  at high energies 
(e.g. 1 TeV electromagnetic showers versus 2 TeV hadronic showers), 
becomes significantly larger at very low energies
below 10 GeV   
(e.g. 2 GeV $\gamma$-ray events  versus 30 GeV hadronic showers). 
%
\begin{figure}[htbp]
\begin{center}
\includegraphics[width=0.48\linewidth]{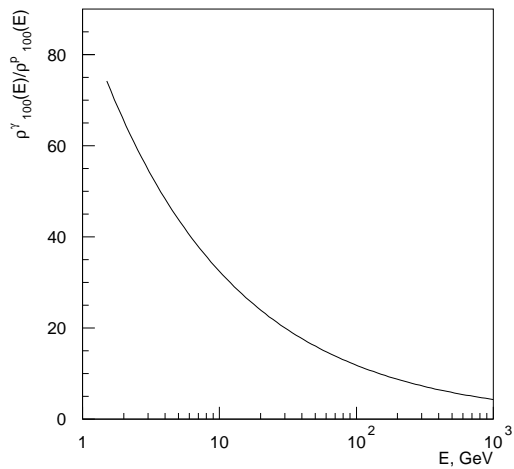}
\caption{The ratio of densities of the 
Cherenkov light at the  altitude of 5 km a.s.l.  
produced by $\gamma$-rays and protons at 
100 m distance from the shower core.  
}
\end{center} 
\end{figure}

\begin{figure}[htbp]
\begin{center}
\includegraphics[width=0.48\linewidth]{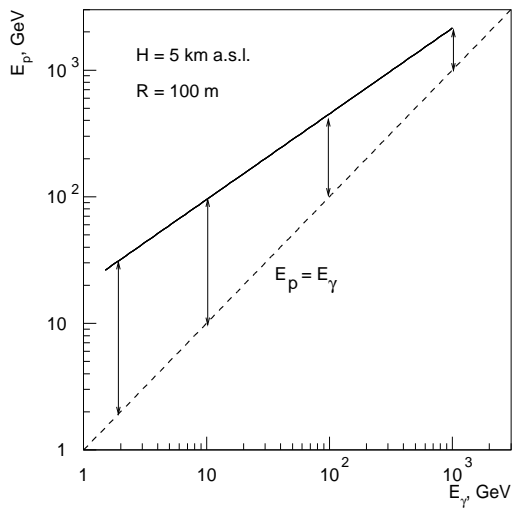}
\caption{The relation 
between energies of $\gamma$-rays 
$E_\gamma$ and protons $E_{\rm p}$
defined from the condition of production of the
same amount of the Cherenkov light 
at 100 m from the shower core (solid line). 
The dashed line corresponds to $E_\gamma=E_{\rm p}$.
The vertical lines at several energies of $\gamma$-rays  
illustrate the gaps between 
$E_\gamma$ and $E_{\rm p}$.
}
\end{center} 
\end{figure}

\subsection{Field of view and pixel size}

One of the principal parameters  of the imaging camera  
is its field of view (FoV). At TeV energies 
the image centroids of showers detected at distances $R \sim 100 \, \rm m$
or beyond are shifted in the focal plane from the center of the camera by
$\approx 1^{\circ}$. Therefore an inner region with diameter $ \sim 3^{\circ}$ 
provides high detection efficiency for $\gamma$-rays, 
and can  be  treated as optimal zone for the hardware trigger.  
On the other hand, the IACT technique requires 
that the camera should be larger 
than the trigger zone by about one degree,  
in order to avoid a  distortion of the Cherenkov images 
because of a  limited FoV;  for lower   
$\gamma$-ray energies  the shift of the image centroid 
decreases \cite{array}.  This
effect becomes significant especially at very low energies.
\begin{figure}[t]
\begin{center}
\includegraphics[width=0.7\linewidth]{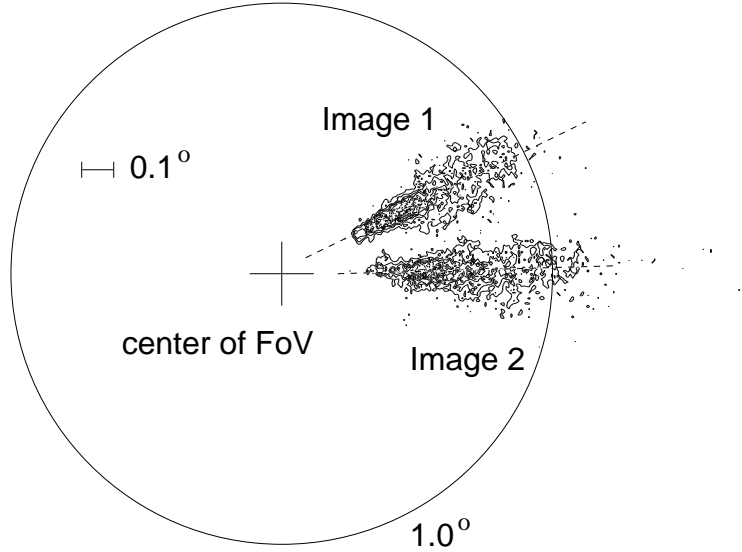}
\caption{The Cherenkov images  
in the focal plane of the camera from 10 GeV 
$\gamma$-ray showers with impact parameters of 
100 m (Image 1) and 125 m (Image 2).}
\end{center} 
\end{figure}
In particular at $E=10 \, \rm GeV$
the shift does not exceed $0.5^{\circ}$ (Fig.~7). 
Thus in the energy range  $E \ll 100 \, \rm GeV$, the camera 
can  be more compact compared with the conventional
(FoV$\geq 4^{\circ}$) cameras designed for  high energies. 
In fact,  a FoV$\approx 3^{\circ}$ seems a reasonable
for point-like or even moderately extended 
($\leq 0.5^{\circ}$) $\gamma$-ray sources. This is an important 
circumstance  which keeps the number of the channels of the imager 
within  reasonable ($\leq 10^3$) limits, especially  
taking into account that at such low energies  
the pixel size must  be small,  
$\approx 0.1^{\circ}$ or less,   dictated by two 
the equally important conditions:   effective suppression of 
the N.S.B.,  and high image quality (see Sec.~2).

\subsection{Trigger integration gate}

Truncation of the Cherenkov pulse integration gate is another 
effective way  to suppress  the N.S.B. Ideally, 
the integration gate should be comparable with the  
duration of the time impulse of the Cherenkov light  
which  for 10 GeV $\gamma$-rays, detected at 5 km a.s.l. 
altitude,  is less than 5 ns (Fig.~8). The compression 
of the integration gate from conventional 20 ns 
to 5 ns reduces the average noise level - the number of photoelectrons 
produced by the N.S.B. -  by a factor of 4, 
and correspondingly lowers  the energy threshold  significantly.

\begin{figure}[htbp]
\begin{center}
\includegraphics[width=0.6\linewidth]{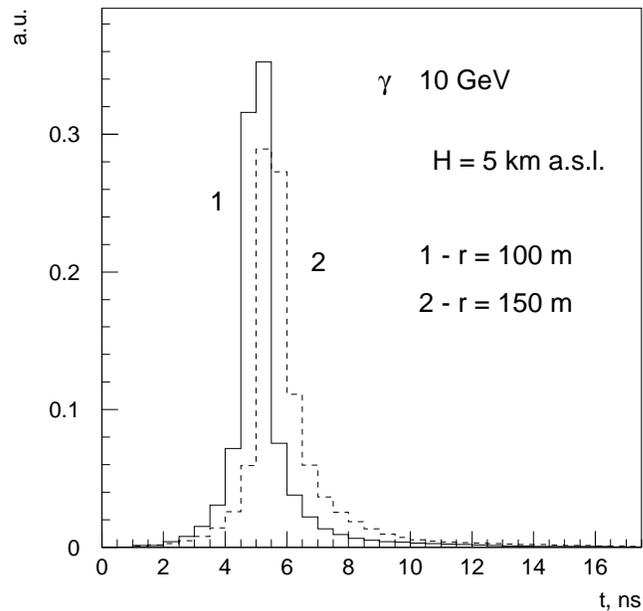}
\caption{Time distribution of the Cherenkov light pulses
from a 10 GeV $\gamma$-ray shower at 100 m and 150 m 
distances from the shower core.}
\end{center} 
\end{figure}
Note that both techniques,   
{\em fast timing} and
{\em small pixel size}, have
already been successfully employed by the CAT collaboration  \cite{cat}.
A  further suppression of the N.S.B.
can be effectively achieved in  the stereoscopic approach
by requiring simultaneous detection of showers by two or more
telescopes. This technique  has been convincingly demonstrated 
by the HEGRA collaboration \cite{Daum,hegra_MC}.
Below we will show that,  even for telescopes with an aperture as large as
$50 \, \rm m^2$, the combination of all three techniques - 
{\em small pixel size} ($\sim 0.1^{\circ}$), 
{\em fast timing} ($\leq 5 \rm \ ns$),
and {\em stereoscopy} -  allows effective operation of the 
IACT system in the energy regime as low as several GeV.   
 
\subsection{Effect of geomagnetic field}

An important issue concerning the quality 
of Cherenkov images at very low energies 
%
\begin{figure}[htbp]
\begin{center}
\includegraphics[width=0.6\linewidth]{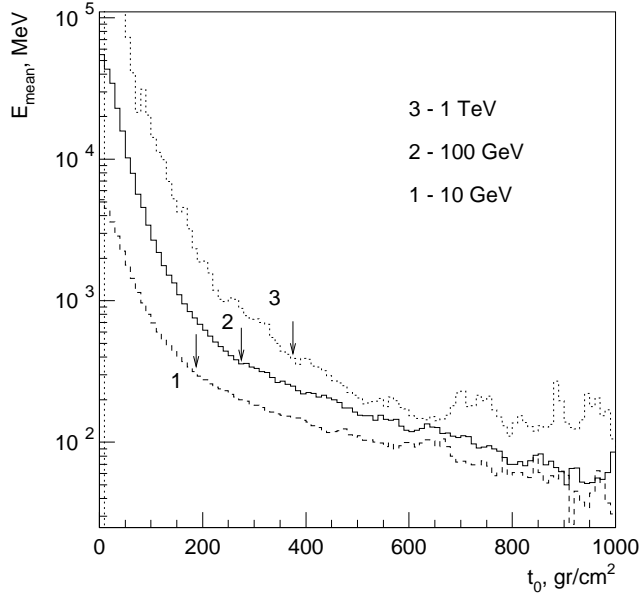}
\caption{The average energy of the secondary electrons
contributing to the bulk of the Cherenkov light of 
10 GeV, 100 GeV, and 1 TeV $\gamma$-ray showers
as a function of the atmospheric depth 
along the shower axis. The arrows indicate  
the  positions of the shower maximum for these energies.
One can see that the average energy of electrons in the 
shower maximum, $E_{\rm mean} \approx 300 - 400 \, \rm MeV$,
is almost independent of the energy of primary $\gamma$-rays.
}  
\end{center} 
\end{figure}
\begin{figure}[htbp]
\begin{center}
\includegraphics[width=0.4\linewidth]{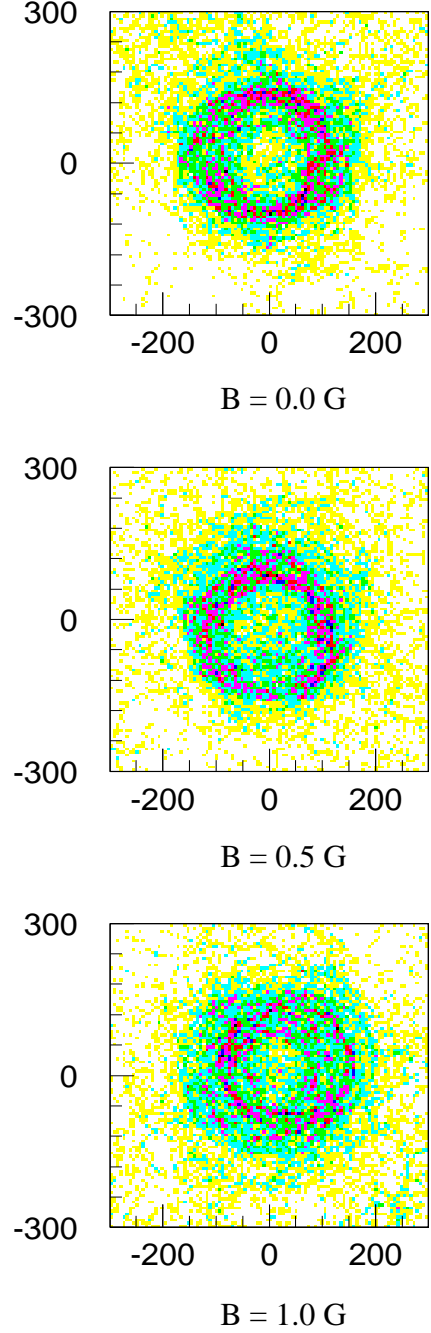}
\caption{The density of Cherenkov light at the observation level
from 10 GeV $\gamma$-ray showers for 3 different values of
the perpendicular component of the 
geomagnetic field.} 
\end{center} 
\end{figure}
%
is connected with the deflection of the secondary 
(cascade) electrons in the geomagnetic field (see e.g. Ref. \cite{durham}). 
Generally, our results agree well with the conclusion of 
%
\begin{figure}[htbp]
\begin{center}
\includegraphics[width=0.6\linewidth]{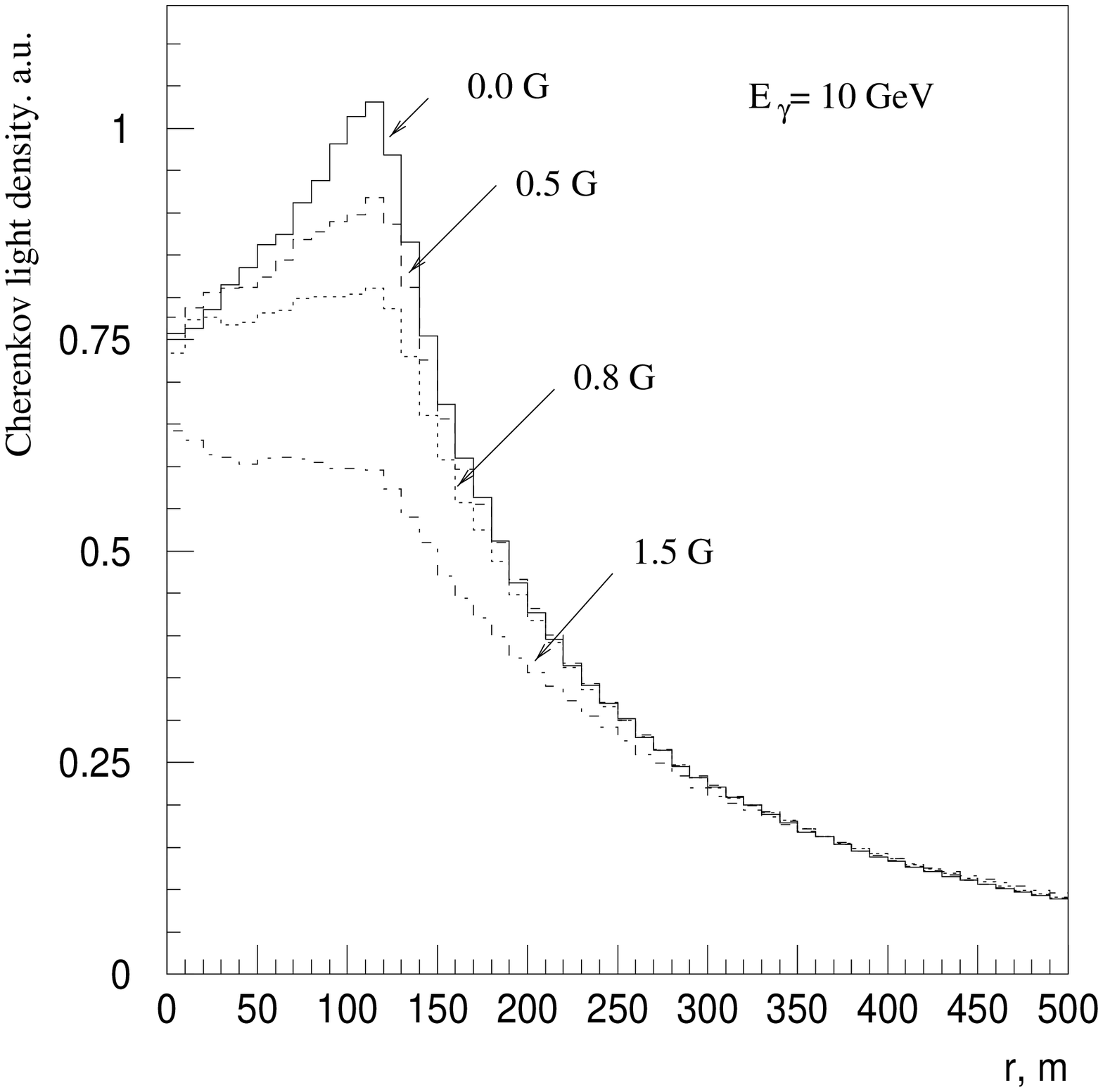}
\caption{The effect of the geomagnetic field on the
lateral distribution of 10 GeV $\gamma$-ray showers
for different values of the perpendicular component
of the geomagnetic field.} 
\end{center} 
\end{figure}
%
Patterson and Hillas \cite{B_hillas} that 
down to $\gamma$-ray energies of 100 GeV 
the geomagnetic field has a rather small effect on the lateral 
distribution of the Cherenkov radiation as long as the 
perpendicular component of the magnetic field  $B_{\perp}$  is less than 0.3 G.
One cannot {\em a priory} exclude, however,  
that at very low energies the effect of the geomagnetic field 
is  stronger. 
The main parameter characterizing the 
effect of the geomagnetic field is the average energy $\overline{E_{\rm e}}$ 
of  the electrons  radiating Cherenkov light at the 
shower maximum, rather than the
energy of primary particles. In Fig.~9 we 
compare the parameter
$\overline{E_{\rm e}}$ for 
showers produced by primary $\gamma$-rays at  
three  energies - 10 GeV, 100 GeV, and 1 TeV.
It is seen that, at any fixed depth,  the average
energy of electrons producing Cherenkov light is
larger for cascades induced by higher energy $\gamma$-rays. However,
the average electron energies become quite 
similar when we compare $\overline{E_{\rm e}}$ at 
the depths corresponding to the shower maxima for 
the given energy of the primary $\gamma$-ray photon. 
This interesting effect is explained by the fact that 
at low primary energies  the maximum of an electromagnetic 
shower occurs at high altitudes where the density is low, 
and therefore the electrons  need to be more energetic
to produce Cherenkov light. In practice this implies that
lowering the energy of $\gamma$-rays should not 
result in  significant amplification of the  geomagnetic effect.
In Fig.~10 the density of the Cherenkov
light from 10 GeV showers at the observation level 
are shown for three  different values  of the perpendicular component 
of the geomagnetic field - $B_{\perp}$=0, 0.5 G, 1 G. The split 
of the patterns (azimuthal asymmetry) 
produced by electrons and positrons due to 
their deflections in opposite  directions becomes noticeable only
for a very large perpendicular component of the field, 
$B_{\perp}$=1 G. More quantitatively, the effect of the geomagnetic field 
on the lateral distribution of Cherenkov photons from 10 GeV 
$\gamma$-rays is shown in Fig.~11. It is seen that for a reasonable
field $B_{\perp} \leq 0.5 \, \rm G$, 
the effect is less than  15 per cent.

\section{Detection areas} 

For a given telescope configuration and    
arrangement of the IACT array, the energy threshold 
and the effective detection area of primary $\gamma$-rays
are determined by the lateral and angular distributions 
of the Cherenkov light and the hardware trigger conditions.
%
\begin{figure}[htbp]
\begin{center}
\includegraphics[width=0.6\linewidth]{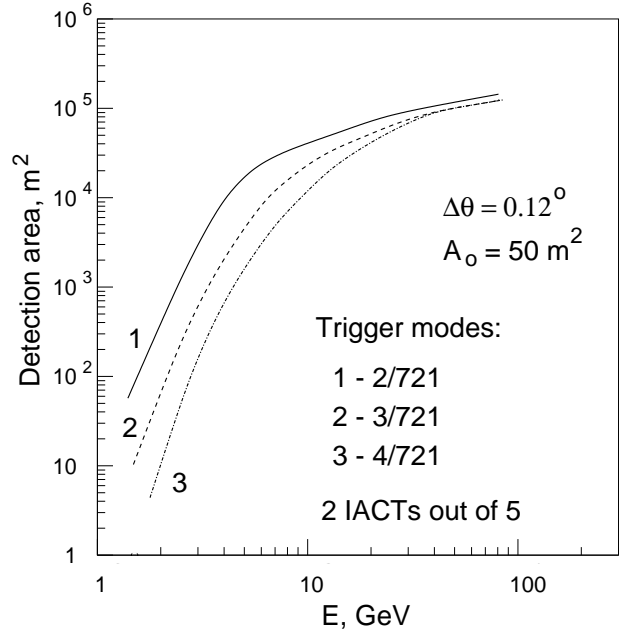}
\caption{Effective collection areas for $\gamma$-ray showers 
calculated for different trigger conditions.}
\end{center} 
\end{figure}
%
The configuration of the IACT array assumed in this paper is the 
following: the system consisting  of 5 IACTs is installed 
at the altitude $H=5 \, \rm km$ a.s.l.,  four telescopes are 
located at the corners, and one in the center of a square 
with a linear size $d=100 \, \rm m$;  each telescope has an
aperture $S=50 \, \rm m^2$, and is equipped with a $n=721$-channel
camera of individual pixel size $0.12^{\circ}$. This comprises 
an effective field of view FoV$=3.2^{\circ}$. 
The hardware trigger is organized in a way that requires 
(1) signals above some critical threshold $q_0$
in $m$ adjacent pixels (``$m/n \geq q_0$'') in each 
individual telescope (the ``local trigger''); (2) detection 
of a shower by at least 2 telescopes (the ``system trigger'').     
For calculations of the threshold $q_0$ we assume a 5 ns
trigger gate, and require that the accidental rate
caused by the N.S.B. is less than 1 Hz, i.e.  
less than 1 per cent of the detection rate of cosmic ray electrons 
(see below).     

\begin{figure}[htbp]
\begin{center}
\includegraphics[width=0.6\linewidth]{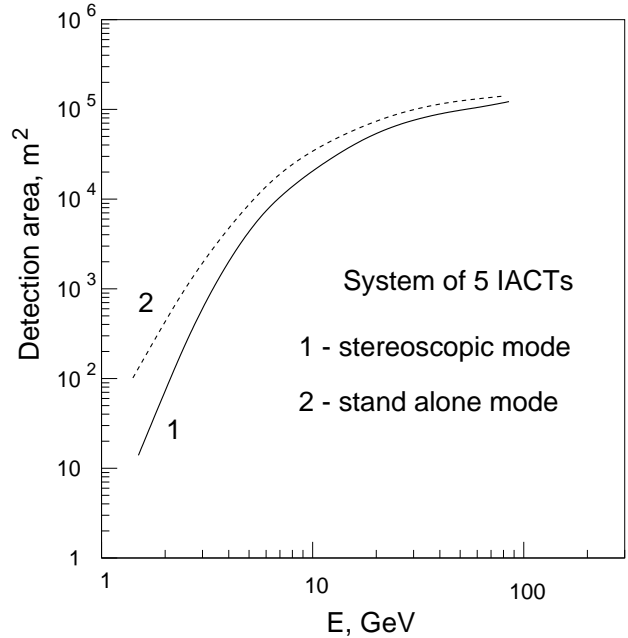}
\caption{Effective detection areas for $\gamma$-ray showers
for 5 IACTs working in the  stereoscopic mode 
(at least two telescopes in coincidence) with 
trigger condition $2/721 \geq q_0$ with  $q_0=$8 ph.-e., and 
in a  stand alone mode  with  $q_0=$11 ph.-e.}  
\end{center} 
\end{figure}

In Fig.~12 we show the detection areas for $\gamma$-rays from
1.5 GeV to 100 GeV calculated for 3  local  trigger conditions:
$2/721 \geq$ 6 ph.-e., $3/721 \geq$ 5 ph.-e.
$4/721 \geq$ 5 ph.-e., assuming  
a standard N.S.B. flux  
$F_{N.S.B.} \simeq 1.5 \times 10^{12} \, \rm ph/m^2 sr \ s$  
(see e.g. \cite{Cawley}).  
It is seen that 
for a chosen pixel size of $0.12^{\circ}$,
a multiplicity greater than two ($m > 2$) actually 
reduces the detection area, even though the requirement 
of higher multiplicity allows a lower trigger threshold $q_0$. 
Therefore all  calculations presented below
were performed  for $m=2$. 

In Fig.~13 the detection  area of the IACT array
for a somewhat  higher threshold,  $q_0=8$ ph.-e.,  is shown in order to
demonstrate how sensitive  the detection area is to the choice 
of  $q_0$, especially at low, sub-10 GeV energies (compare 
curves 1 in Fig.~12 and Fig.~13). 
For comparison, we show also the overall detection area of 
5 telescopes if they would be located at large,
$\gg 100 \, \rm m$ distances from each other 
and operating independently in a  stand alone  mode, 
i.e. being not integrated in the system trigger.
It is seen that at energies $E \geq 3 \, \rm GeV$ the array
of independent telescopes has a  larger detection area, 
but  the difference 
compared with the array operating in  the  stereoscopic 
mode at energies  $E \geq 5 \, \rm GeV$ does not exceed 
a factor of two. On the other hand, the analysis of 
images obtained in the stereoscopic mode 
provides  much better suppression of the 
background caused by the cosmic ray protons
and electrons  which effectively results in   
a significant improvement of not only the quality of the 
data, but also of the flux sensitivity of the instrument.

At energies below 100 GeV, the cosmic ray background 
detected by the  system of IACTs operating in the 
stereoscopic mode is essentially dominated by electromagnetic 
showers produced by cosmic ray electrons. 
This  is seen in Fig.~14 where the detection rates
of both cosmic ray protons and electrons  are also shown. 
The strong dominance of the electronic component 
is explained by the combination of several effects, in particular 
{\it (i)} the large, up to a factor of 10,  difference between
the energies of electrons and protons producing the same amount of
Cherenkov light, {\it (ii)} the high altitude of observations,
{\it (iii)} the compact Cherenkov images of electromagnetic showers 
compared with hadronic showers,
{\it (ii)} the noticeable  increase (approximately  $\propto E^{-0.5}$) 
of the electron-to-proton ratio 
of cosmic rays down to  $E \sim 10 \, \rm GeV$.  

In Fig.14 we show also the detection rate of $\gamma$-rays from  a
point source with a power-law spectrum 
${\rm d} J_\gamma/{\rm d} E \sim E^{-2.5}$, and integral flux 
$J_\gamma(\geq 1 \, \rm GeV)=3 \times 10^{-7} \, \rm ph/cm^2 s$.
The latter  is somewhat larger than the fluxes  of most of the EGRET sources 
\cite{Vela}. This implies that the $\gamma$-ray detection rate shown in Fig.~14
should be considered as an  upper limit for ``standard'' 
EGRET sources. This curve still lies significantly below the rate of detection 
of cosmic ray  electrons. However, for a  point-like 
source the  electron background can  be reduced significantly 
if we select showers  arriving  from the direction of the 
$\gamma$-ray source.

\begin{figure}[htbp]
\begin{center}
\includegraphics[width=0.6\linewidth]{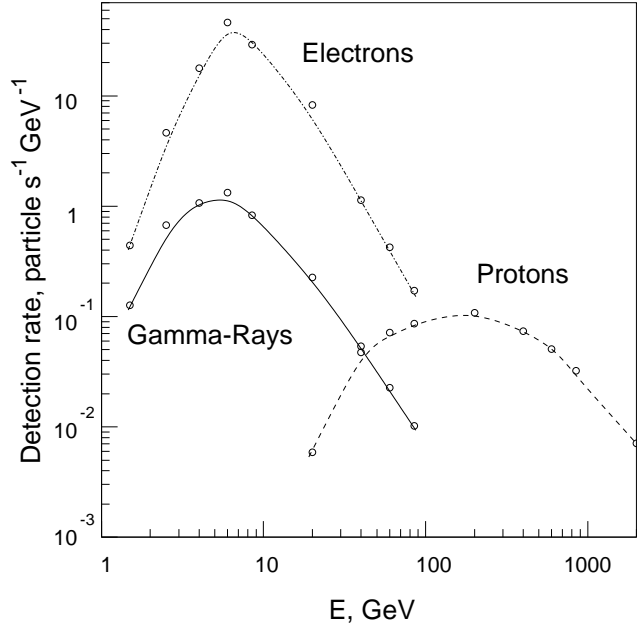}
\caption{Differential detection rates of showers produced
by $\gamma$-rays and cosmic ray protons and electrons.
For $\gamma$-rays we assume a power-law spectrum with photon index 
2.5, and integral flux 
$J_\gamma(E \geq 1 \, \rm GeV)=3 \times 10^{-7} \, \rm ph/cm^2 s$.}  
\end{center} 
\end{figure}

The detection rates of electromagnetic showers within the 
detector's energy-dependent Point Spread Function (PSF), 
given by the angle  
$\phi$, which is  the half-angle of the cone around the 
source direction containing $67 \%$ of events at energy $E$, 
are determined as follows:
\begin{equation}
R_\gamma=\frac{{\rm d}J}{{\rm d}E} \ A_{\rm eff}(E) \ \kappa_\gamma \ , 
\end{equation}    
and
\begin{equation}
R_{\rm e}=\frac{{\rm d}J_{\rm e}}{{\rm d}E {\rm d}\Omega} 
\ A_{\rm eff}(E) \Omega \ , 
\end{equation}    
for $\gamma$-rays  and cosmic ray electrons, respectively,
where ${\rm d}J/{\rm d}E$ is the differential flux of $\gamma$-rays
from a point source, ${\rm d}J_{\rm e}/{\rm d}E {\rm d}\Omega$
is the differential flux of cosmic ray electrons per solid angle, $A_{\rm eff}$ is the 
detection area for  electromagnetic showers, $\Omega=2 \pi \ (1- \cos \phi)$,
and $\kappa_\gamma=0.67$ is (by definition) the $\gamma$-ray acceptance.

\begin{figure}[htbp]
\begin{center}
\includegraphics[width=0.6\linewidth]{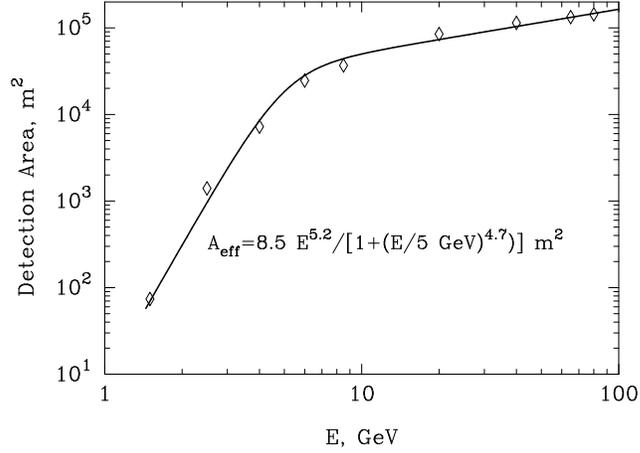}
\caption{Detection area of the 5 GeV IACT array
calculated for the trigger threshold $q_0=$6 ph.-e.
The solid curve corresponds to the analytical
fit represented by Eq.(3).}  
\end{center} 
\end{figure}

\begin{figure}[htbp]
\begin{center}
\includegraphics[width=0.6\linewidth]{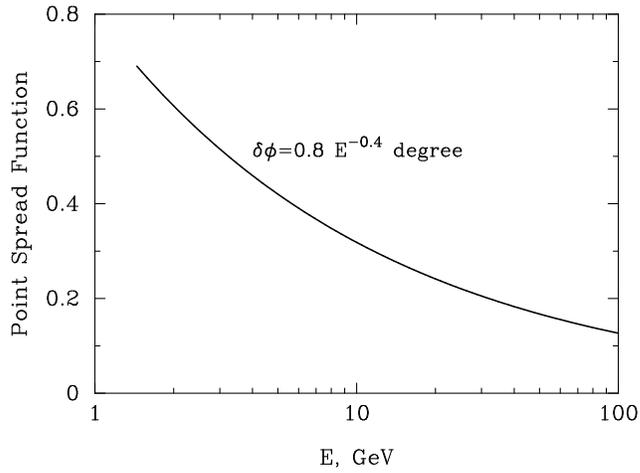}
\caption{Point spread function of the  5 GeV IACT array.}
\end{center} 
\end{figure}

For the chosen configuration of the IACT array, the Monte-Carlo
calculations of the effective detection area $A_{\rm eff}$ 
(Fig.~15) in the  interval from 1.5 GeV to 100 GeV
can  be presented  in  the following form  
\begin{equation} 
A_{\rm eff}=8.5 E^{5.2} [1+(E/5 \, \rm GeV)^{4.7}]^{-1} \, \rm m^2 \, , 
\end{equation}
where E is the energy of a $\gamma$-ray or electron in units of GeV.

\begin{figure}[htbp]
\begin{center}
\includegraphics[width=0.6\linewidth]{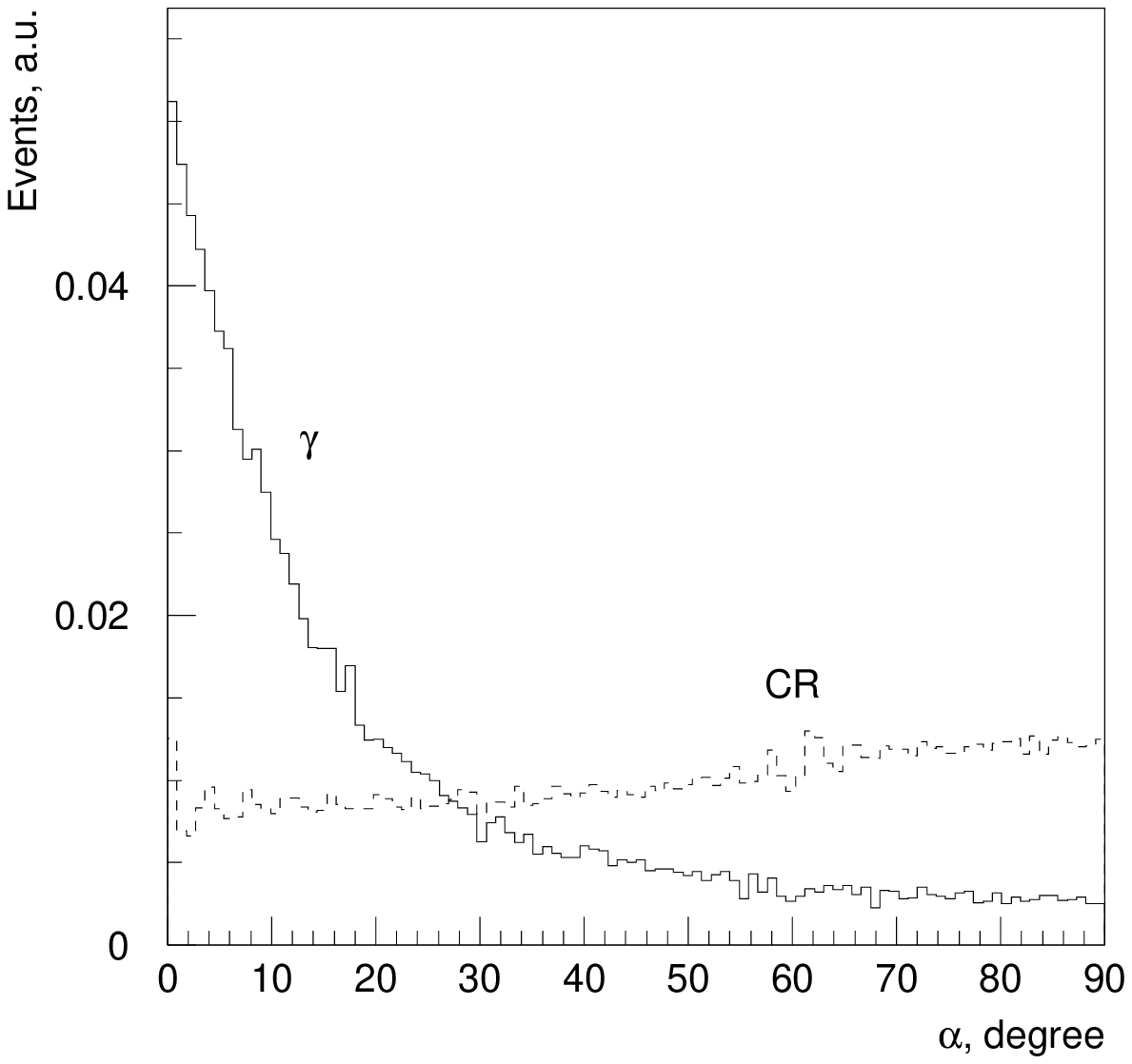}
\caption{The {\sc alpha} distributions for a single telescope.}
\end{center} 
\end{figure}

\vspace{3mm}

\begin{table}[htbp]
\caption{Efficiency of the {\sc alpha}-cut for a single ``low energy''
telescope.}
\bigskip
\begin{center}
\begin{tabular}{llll}
Cut, deg & $\kappa_\gamma$ & $\kappa_{CR}$ & Q-factor \\ \hline
5        & 0.25            & 0.0504        & 1.11 \\
10       & 0.42            & 0.0972        & 1.35 \\
15       & 0.54            & 0.1439        & 1.42 \\
20       & 0.62            & 0.1907        & 1.42 \\
25       & 0.69            & 0.2348        & 1.42 \\
30       & 0.73            & 0.2868        & 1.36 \\ \hline
\end{tabular}
\end{center}
\end{table}

This presentation shows a strong energy-dependence
of the detection area at energies below 10 GeV, 
$A_{\rm eff}(E) \propto E^{5.2}$, but at higher energies 
it gradually turns into a slow increase with energy, 
$A_{\rm eff}(E) \propto E^{0.5}$  
Close to 1 TeV  the detection area actually becomes constant, in essence 
because of the  limited field of the view ($\approx 3^{\circ}$)
of the camera.  

The stereoscopic approach allows the determination of the 
arrival direction of primary $\gamma$-rays on an event-by-event
basis. The determination of the 
arrival direction of primary $\gamma$-rays is described in
ref.\cite{array}, and its practical 
implementation in the case of the HEGRA IACT system
can  be found in ref. \cite{hegra_MC,wh_utah}. 
In Fig.~16 we show the Monte-Carlo  calculations of the PSF   
which can be approximated in the simple form 
\begin{equation} 
\phi=0.8 (E/1 \, \rm GeV)^{-0.4} \, \rm degree \, .
\end{equation}

Despite the small pixel size $\sim 0.1^{\circ}$, 
the angular resolution  of the  5 GeV IACT array   
at energies $E \ll 100 \, \rm GeV$ is 
significantly poorer  than the  resolution of  
``100 GeV'' threshold instruments,
which could be as good  as $0.1^{\circ}$,  even for a 
larger pixel-size of about $0.25^{\circ}$ \cite{array}.
Since for both categories of  instruments the number 
of photoelectrons, or the so-called size of the image,    
from a detected $\gamma$-ray photon 
are comparable ($\sim 100 (E/E_{\rm th})$ 
photoelectrons), the lower performance of the IACT technique 
is rather an intrinsic feature of Cherenkov images
at very  low energies (the images are 
less elongated and less regular).
This effect can  be seen also in the so-called {\sc alpha}-distribution 
of images in a single telescope, where the {\sc alpha} 
parameter is indicative of the orientation of the image 
in the camera (see e.g. ref.\cite{Fegan}):  
while at TeV energies most  of $\gamma$-rays from a 
point source have an angle {\sc alpha} less than 
5-8 deg  \cite{Fegan},   the {\sc alpha}  distribution is significantly  broader
in the energy region around 
10 GeV (see Fig.~17). 
The suppression of the cosmic ray background at such low energies becomes 
correspondingly less effective.  The characteristic values of the acceptance of 
the isotropic cosmic ray showers $\kappa_{\rm CR}$, 
and the point-source $\gamma$-ray showers  $\kappa_\gamma$,
as well as the so-called Q-factor, 
$Q=\kappa_\gamma/\kappa_{\rm CR}^{1/2}$, which characterizes the 
improvement of the signal-to-noise ratio after application
of the image cuts, are shown in Table 1.
It is seen that the best improvement of  the signal-to-noise
ratio is achieved for an  {\sc alpha} cut at $\sim 20-25^{\circ}$
which gives a rather modest Q-factor, $Q \simeq 1.4$. In contrast,
at 10 GeV the stereoscopic measurements allow the determination 
of the shower direction with an accuracy of $0.3^{\circ}$ (Fig.~16). 
Therefore the $\gamma$-ray signal could be improved
by a factor of $Q=\kappa_\gamma/\kappa_{\rm CR}^{1/2}=
\kappa_\gamma (\Psi/2 \phi) \approx 3.35$,
where the efficiency of the rejection showers from
cosmic ray electrons  
$\kappa_{\rm CR} \approx (2 \ \phi/\Psi)^2$
is determined by the FoV of the imager,
$\Psi \simeq 3^{\circ}$, and the angular resolution 
$\phi \simeq 0.3^{\circ}$. 
Thus, despite 
the smaller   (by a factor of two) detection area of the
array operating in the stereoscopic mode compared 
with the overall area of 5 independent 
IACTs (see Fig.~13), the stereoscopic array
would have at least by a factor of 2 better sensitivity,
even disregarding  other advantages of the stereoscopic 
approach, in particular, the complete removal of the 
hadronic background.

\section{Detection rates and the energy threshold}

The differential detection rates of $\gamma$-rays
from a point source, calculated for the collection area
given by Eq.~(3) and  assuming pure power-law spectrum in the form 
\begin{equation}
\frac{{\rm d}J}{{\rm d}E}=10^{-7} \ (E/1 \, \rm GeV)^{-\alpha} \
\rm cm^{-2} s^{-1} GeV^{-1} \ ,
\end{equation} 
are shown in Fig.~18a. 
Eq.~(5) implies that,
independent of the spectral index $\alpha$,
the differential $\gamma$-ray flux is normalized  
at 1 GeV to  $10^{-7} \ \rm cm^{-2} s^{-1} GeV^{-1}$.
The latter corresponds to the typical flux of  ``standard''  EGRET sources. 

\begin{figure}[htbp]
\begin{center}
\includegraphics[width=0.45\linewidth]{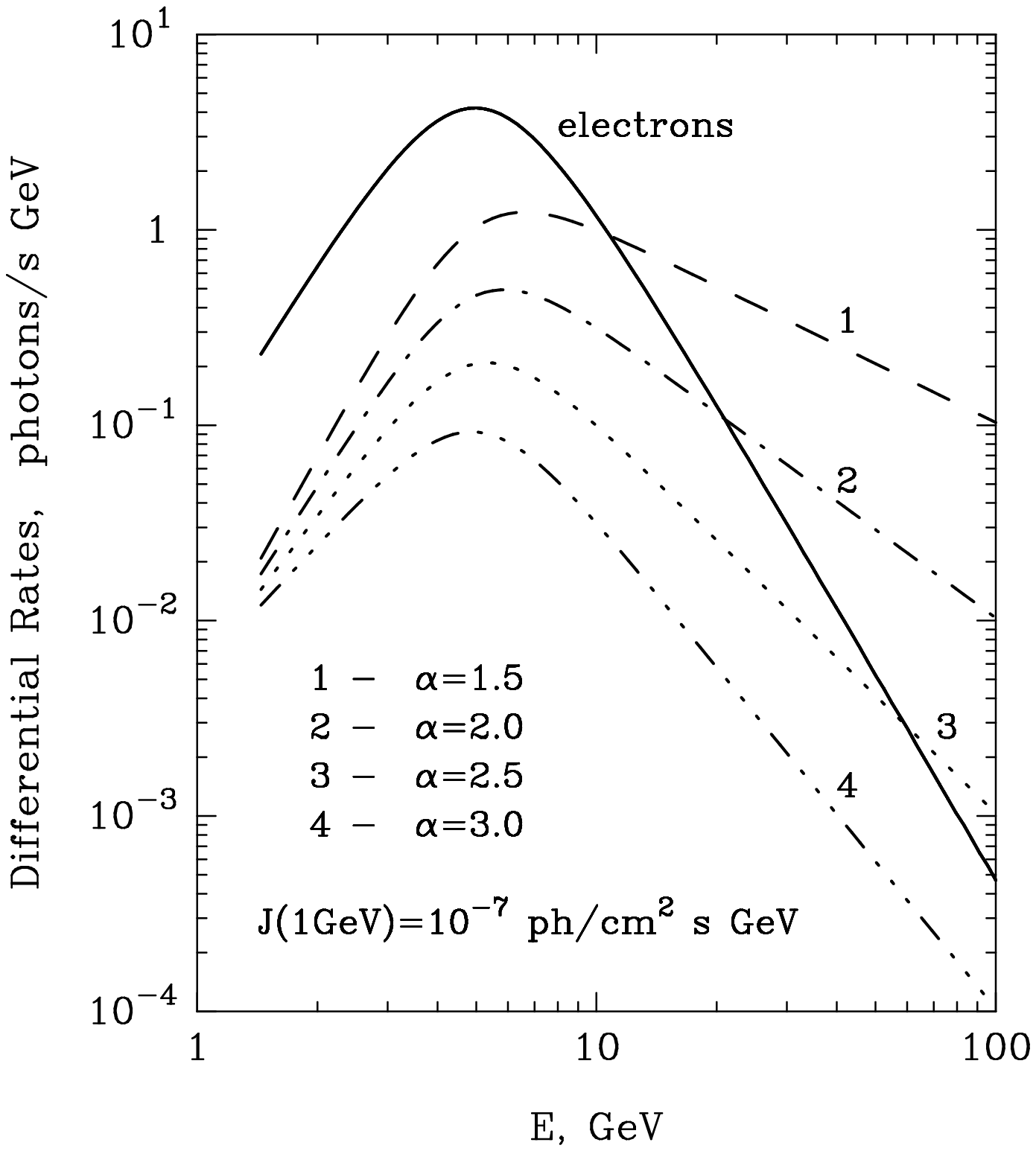}\hspace{5mm}
\includegraphics[width=0.45\linewidth]{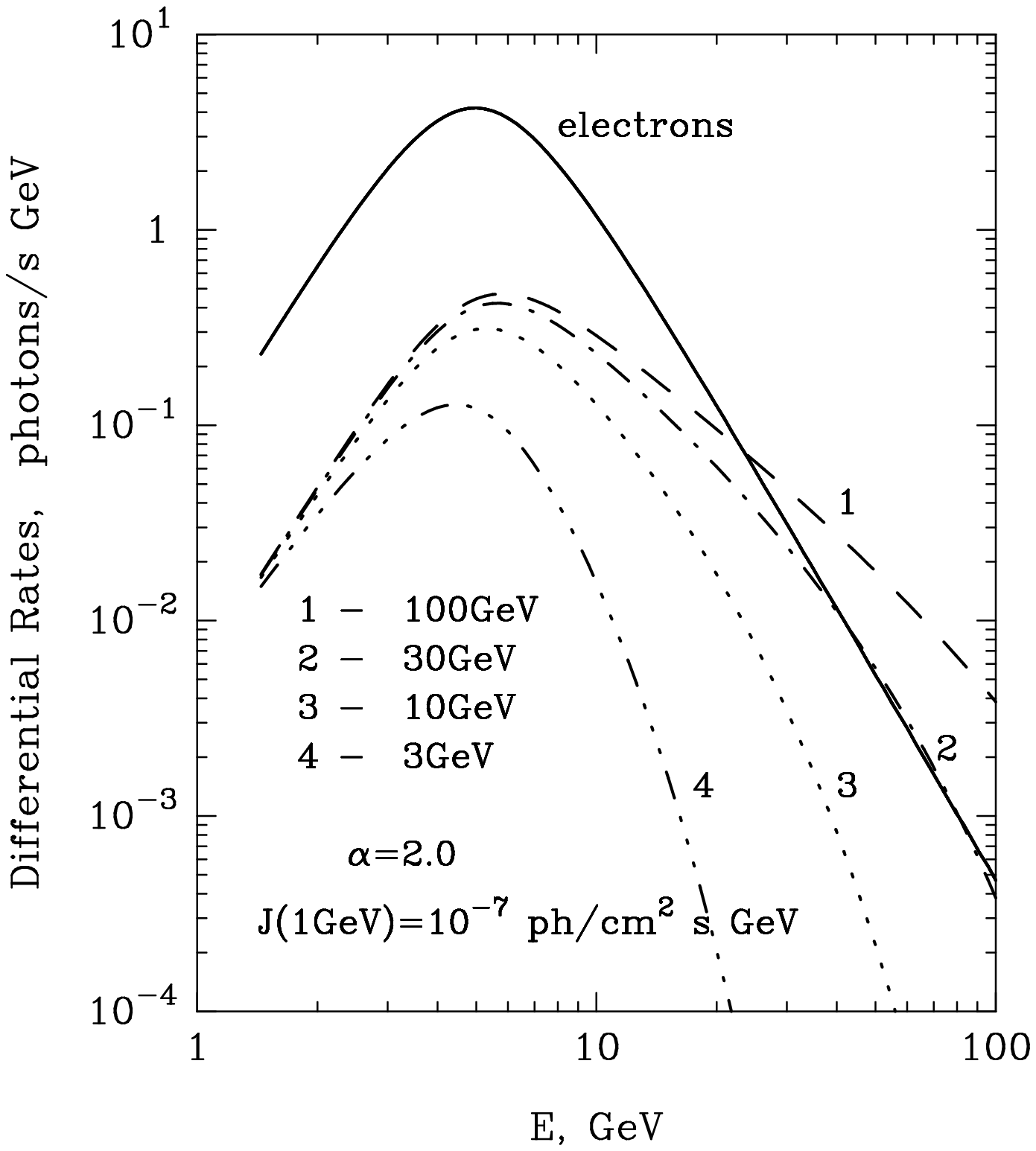}
\caption{Differential detection rates 
of $\gamma$-rays and cosmic-ray electrons within the 
cone determined by the PSF of the IACT array, calculated for 
two types of $\gamma$-ray spectra represented by 
{\bf (a)} Eq.(5) (left) and {\bf (b)} Eq.(6) (right).}
\end{center} 
\end{figure}

In Fig.~18b we show the differential detection rates of $\gamma$-rays 
with  a hard power-law spectrum  with $\alpha=2$  and an  
exponential cutoff at $E_0$:
\begin{equation}
\frac{{\rm d}J}{{\rm d}E} \propto E^{-2} \ \exp(-E/E_0),
\end{equation}   
for 4 different values of $E_0$, and assuming the 
same absolute flux normalization at 1 GeV as in Fig.~18a. 

\begin{figure}[htbp]
\begin{center}
\includegraphics[width=0.6\linewidth]{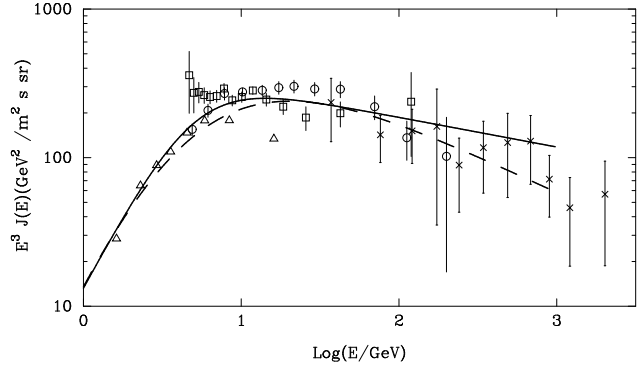}
\caption{The flux of cosmic ray electrons. The compilation
of the experimental fluxes is taken from Ref. \cite{taira,heat}.
The solid line corresponds to the fit represented by Eq.(7).
The dashed line correspond to the function 
${\rm d} J_{\rm e}/{\rm d}E \propto E^{-0.4} [1+(E/3.9 \, \rm GeV)]^{-3.1}$ which provides a better fit to the data at energies above 100 GeV.}  
\end{center} 
\end{figure} 

It is seen from Figs.18a  and 18b that for a 
large variety of $\gamma$-ray spectra the peak of the detection rate
appears in a rather narrow band between 4 and 6 GeV. 
Now,  defining the energy threshold as {\em the energy at which
the differential $\gamma$-ray detection rate reaches to its maximum},
we may conclude that the suggested IACT array has an effective
energy threshold of about 5 GeV.

In Fig.~18a,b we show also the differential detection rate of cosmic ray 
electrons   within the cone limited by the PSF of the instrument 
given by Eq.~(4). The  energy spectrum of cosmic ray electrons
is  shown in Fig.~19.  At energies above 
10 GeV the differential  spectrum is very steep with a power-law index 
$\alpha_{\rm e} \sim 3.2$. 
Below 10 GeV it becomes flatter.
Within the uncertainties of the measured fluxes,  the electron 
spectrum can be approximated  
in the entire region from several GeV to 1 TeV  by the
following function shown by the solid line in Fig.~19 :
\begin{equation} 
\frac{{\rm d}J_{\rm e}}{{\rm d}E {\rm d}\Omega}=1.36 \times 10^{-7} \
E^{-1} [1+(E/5 \, \rm GeV)^{2.2}]^{-1} \, m^{-2} s^{-1} sr^{-1} GeV^{-1} \ . 
\end{equation}
 
In Fig.~20 we show integral detection rates 
$R_\gamma(\geq \rm E)$ for power-law $\gamma$-ray 
spectra represented  by Eq.(5). 
It is seen that for a relatively 
flat $\gamma$-ray spectrum
with $\alpha_\gamma \sim 2$ and for an integral flux above 
1 GeV of $10^{-7} \, \rm ph/cm^2 s$ (this approximately
corresponds to the total, i.e. pulsed plus unpulsed, flux
from the Crab), the detection rate of $\gamma$-rays
from the EGRET sources can  be as high as 6 events per second,
against  the cosmic-ray background rate of about 25 events per sec 
caused  by  cosmic-ray electrons. 
This implies that an observation time of approximately 20-30 sec   
would be sufficient to 
detect a statistically  significant signal from  
such a source. Remarkably, for the brightest persistent
$\gamma$-ray source, the Vela pulsar  
with photon index $\alpha_\gamma \sim 1.7$ from 100 MeV to 10 GeV,
and the integral flux 
$J_\gamma(\geq \rm 1 \ GeV) \approx 1.5 \times 10^{-6} \, \rm ph/cm^2 s$,
the detection rates would exceed 100 events per 1 sec. 
Thus a statistically significant signal from the source 
could be obtained during an observation time less than 1~sec~!  
For the given  normalization of the differential flux at 1 GeV,
$10^{-7} \, \rm ph/cm^2 s \ GeV$, the detection of sources with 
steep $\gamma$-ray spectra would require significantly longer 
exposure. Even so, the time needed
for detection of sources with very steep power-law 
spectra with an index $\alpha_\gamma \sim 3$ 
(like curve 4 in Fig.~18a), or with a sharp, e.g. exponential 
cutoff at  a few GeV (like the curve 4 in Fig.~18b), does  
not significantly exceed 1~h. 

%
\begin{figure}[htbp]
\begin{center}
\includegraphics[width=0.5\linewidth]{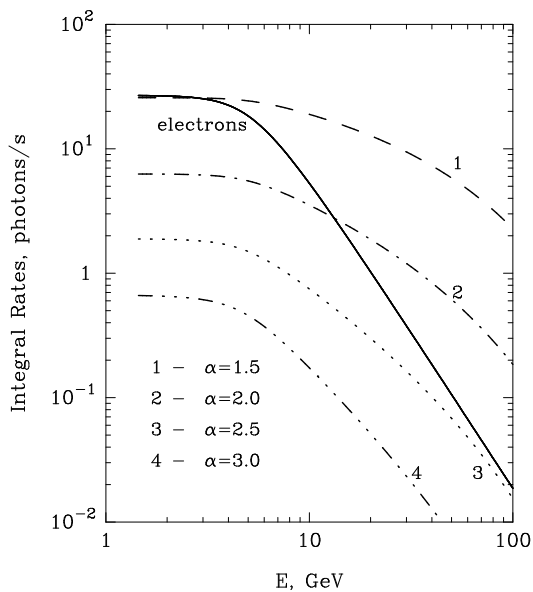}
\caption{Integral detection rates 
of $\gamma$-rays and cosmic-ray electrons within the 
cone determined by the  PSF, calculated for 
$\gamma$-ray spectra represented by 
Eq.(5).}
\end{center} 
\end{figure}

\section{Flux Sensitivity}

The curves in  Fig.~14 correspond to the detection rates  
before the image analysis. Remarkably, even after such
effective rejection of hadronic showers at the 
{\it trigger} level, there still remains room 
for further suppression of the background from
cosmic ray protons and nuclei  by analyzing the 
shapes of the Cherenkov images of the detected showers. 
In Fig.~21 we show the so-called mean-scaled 
{\sc width} parameter  distribution 
of showers which have already passed the hardware trigger condition.
This parameter represents the mean value of the 
{\sc width} parameter measured by all telescopes and 
normalized to the impact distances  and the 
image amplitudes \cite{hegra_MC}.   
It is seen that the distributions of the electromagnetic and 
hadronic showers are rather well separated. The efficiencies of
the acceptance of both type of showers for different 
mean scaled {\sc width} cuts are presented in Table 2. We see that
even very loose cuts at the level of $<w>=1.3$ provide suppression
of the hadronic showers by a factor of 5, while the 
$\gamma$-ray acceptance can  be as high as 90 per cent. 
This implies that after such a loose cut, which practically does 
not reduce the $\gamma$-ray (or CR electron) statistics,  
we may push the detection rates of cosmic ray 
protons and nuclei further down,  
and thus  make 
$\gamma$-ray detection in the entire  energy region 
below 100 GeV essentially free from hadronic background.  
%
\begin{figure}[htbp]
\begin{center}
\includegraphics[width=0.5\linewidth]{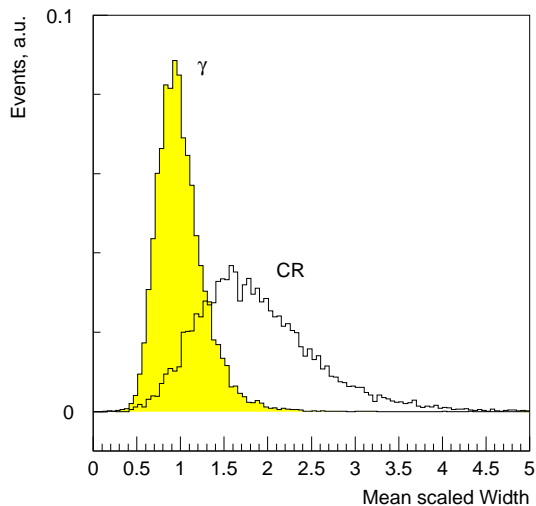}
\caption{Distribution of the mean scaled {\sc width} parameter for
electromagnetic and hadronic showers.}
\end{center} 
\end{figure}
%
\begin{table}[htpb]
\caption{Efficiency of the cosmic ray rejection using the mean 
scaled {\sc width} cut.}
\bigskip 
\begin{center}
\begin{tabular}{lllllll} \hline 
$< \tilde w>$ & 0.8   & 0.9   & 1.0  & 1.1  & 1.2  & 1.3  \\ \hline
$k_\gamma$    & 0.28  & 0.45  & 0.60 & 0.74 & 0.83 & 0.90 \\
$k_{cr}$      & 0.025 & 0.044 & 0.07 & 0.11 & 0.15 & 0.20 \\
Q-factor      & 1.8   & 2.1   & 2.3  & 2.2  & 2.1  & 2.0  \\ \hline 
\end{tabular}
\end{center}
\end{table}
%
This makes the calculations for the {\it differential} 
flux sensitivity of the instrument
straightforward and simple  and,  more importantly,  
there is no  need to specify the spectrum of
primary $\gamma$-rays. Indeed, because the showers produced by 
electrons are very similar to $\gamma$-ray showers\footnote{Actually 
there are some differences.  In particular the primary electrons start 
to produce Cherenkov light earlier, but in this paper we will ignore
these effects.}, the condition of 
detection of a $\gamma$-ray signal  
with statistical significance $m$-sigma in the 
energy interval $[E-\Delta E, E+\Delta E]$,
$N(E) \Delta E=m \sqrt{N_{\rm e} \Delta E}$,
provided that the number of detected $\gamma$-rays 
$N_{\rm min}=J(E) \ 2 \Delta E \ A_{\rm eff}(E) \ T \geq 10$,
gives  the minimum detectable differential flux 
for the observation time $T$:
\begin{equation}
J_{\rm min}(E) \approx \frac{m}{\kappa_\gamma} 
\frac{\phi}{(A_{\rm eff}(E) \ T)^{1/2}} 
\sqrt{\frac{2 \pi J_{\rm e}(E)}{E}}
\end{equation}
where it is assumed that $\Delta E=E/4$, which corresponds to
a rather conservative 25 per cent accuracy of reconstruction
of the energy of the primary electron or $\gamma$-ray photon.  
The results of calculations for the so-called 
spectral energy distribution (SED), $E^2 \ J(E)$, 
based on Eqs.(3),(4) and (7) requiring 3-sigma 
detection ($m=3$) at each energy $E$, are presented in 
Fig.~22 for 2 different observation times, $T=$1 h and 25 h.  
%
\begin{figure}[htbp]
\begin{center}
\includegraphics[width=0.5\linewidth]{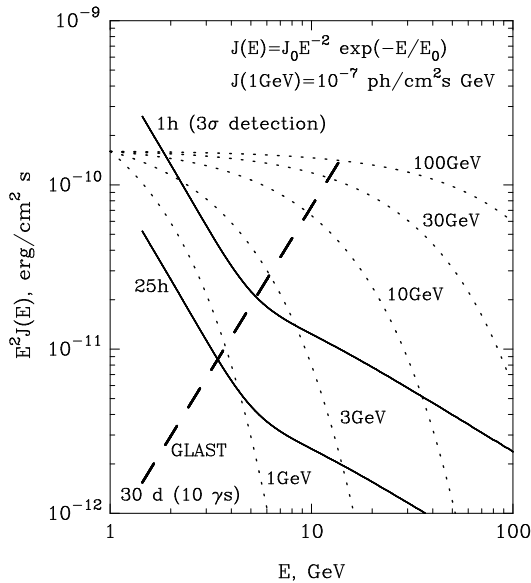}
\caption{Differential flux sensitivities of the IACT array 
for 1 hour and 25 hour
observation times. The expected sensitivity of GLAST for 
30 day continuous observation is shown by the dashed curve.}
\end{center} 
\end{figure}
In the same figure we present also the power-low 
fluxes of $\gamma$-rays  represented by Eq.(6).
It is seen that 1 hour observations by the
IACT array would be sufficient to detect a statistically
significant signal from a ``standard'' EGRET source even 
at the presence of an exponential cutoff in 
the $\gamma$-ray spectrum as low as 3 GeV.  In the case of 
a cutoff at 10 GeV or higher energies, the detection time
($t \propto 1/J^2$) could be reduced to $\leq 1$ min.

\begin{figure}[htbp]
\begin{center}
\includegraphics[width=0.5\linewidth]{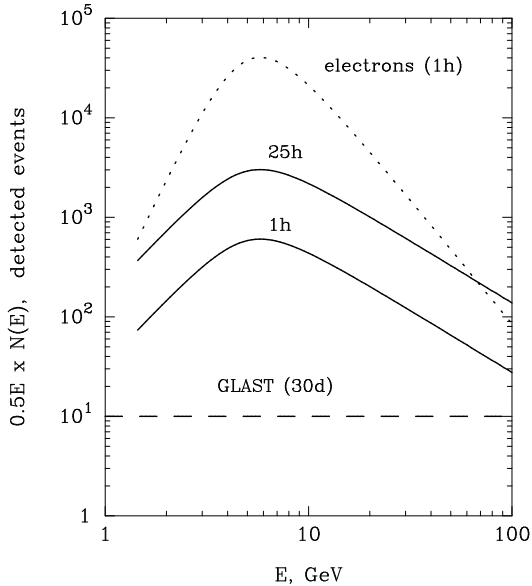}
\caption{Event statistics corresponding to 
the minimum detectable $\gamma$-ray fluxes 
by the IACT array (solid curves) 
and by GLAST (dashed curve) shown in Fig.~20.
The statistics of detected cosmic-ray electrons
by 5 GeV IACT array for 1 h observation 
time is also shown (dotted curve).}
\end{center} 
\end{figure}

For comparison in Fig.~22 we show the expected sensitivity of 
GLAST for an observation time $T=30 \, \rm days$. Since at energies above 
several GeV GLAST will operate at almost background free conditions
(for point-like sources), the flux sensitivity is determined by
the photon statistics, $N_{\rm min}=10$ (see e.g. \cite{Bloom,GehMich}). 
Note that at energies above 5-10  GeV,  1 hour observation time
by the IACT array could provide better sensitivity that the 
minimum detectable fluxes achievable by GLAST during 1 month
of continuous observations. Moreover, even very short
observations by the IACT array can give 
unusually rich (for $\gamma$-ray astronomical standards)
photon statistics over the whole energy region from few GeV to 
100 GeV; the number of detected $\gamma$-rays exceeds
100 at each energy interval $E \pm E/4$ (see Fig.~23).
This would guarantee an appropriate $\gamma$-ray spectroscopy
with  energy resolution of about 20-25 per cent below 
10 GeV, and better than  15 per cent at higher energies.

The flux sensitivity shown in Fig.~22 is obtained under 
a very robust  and to a large extent non-standard condition 
which requires detection of a signal with at least 3-sigma
significance  in  {\em each} energy band with width 
$E/2$ centered on $E$, provided that the number of detected 
$\gamma$-rays in this band exceeds 10.  Note  that this
definition of sensitivity does not require any  
knowledge about the shape of the spectrum of the 
primary $\gamma$-rays.
%
\begin{figure}[htbp]
\begin{center}
\includegraphics[width=0.5\linewidth]{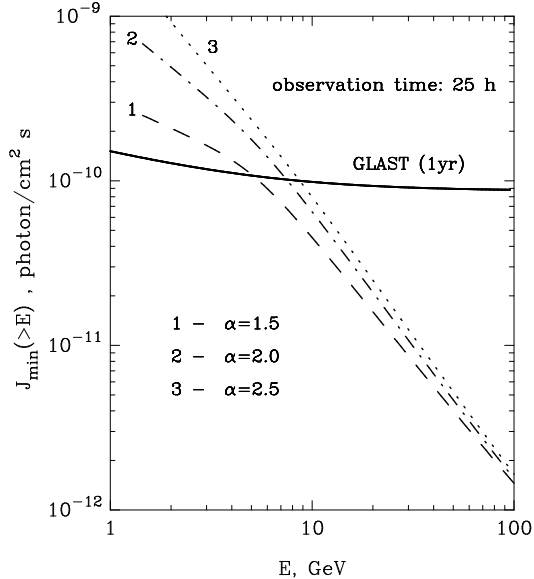}
\caption{Integral  flux sensitivities 
of the IACT  array for 25 h
observation time, assuming power-law 
$\gamma$-ray spectra with photon indices
$\alpha_\gamma=1.5$, 2, and 2.5.  
The sensitivity of GLAST for 
1 year continuous observation time is also shown (solid  curve).}
\end{center} 
\end{figure}
In Fig.24 we present the flux sensitivities determined in a more
traditional way, namely requiring 
5-sigma detection of $\gamma$-rays 
above the given energy, $J_{\rm min}(\geq \rm E)$.
This definition of the integral flux sensitivity 
obviously requires an assumption about the shape of the energy spectrum.
The curves shown in Fig.~24 are calculated for 
$T=25 \, \rm h$ observation time, assuming power-law 
spectra of $\gamma$-rays with photon indices 
$\alpha_\gamma=$1.5 (dotted curve), 2 (dot-dashed curve), and 
2.5 (dashed curve). The expected GLAST sensitivity
shown by the solid curve corresponds to  
1 year of continuous observations of the source.

\section{Discussion}

Results presented in this paper show that a stereoscopic
array of large, 20 m diameter class  imaging Cherenkov telescopes 
installed  at very high mountain altitudes could  effectively 
enter into the domain of satellite-borne  $\gamma$-ray astronomy. 
A {\bf 5} GeV energy threshold array of IACTs {\bf at} {\bf 5} 
km a.s.l. - hereafter 5@5 - could  provide a deeper probe of $\gamma$-ray
sources compared with GLAST - the most powerful current satellite-borne 
$\gamma$-ray project. However, the scientific goals of these instruments are 
essentially different. While GLAST with its almost 
$2 \pi$ steradian  field of view can  provide very effective 
{\it simultaneous}  monitoring of a  very 
large number (hundreds or even thousands)  quasi-stable 
$\gamma$-ray sources, as well as  a study 
the galactic and extragalactic components
of the diffuse $\gamma$-ray background radiation, 
5@5  has an obvious 
advantage for the search and study of highly variable 
or transient $\gamma$-ray sources. The flux sensitivity 
of this instrument at 5 GeV of about 
$2 \times 10^{-11} \, \rm erg/cm^2 s$ (see Fig.~22)
would allow the detection of  any $\gamma$-ray flare 
with apparent luminosity  
$2 \times 10^{39} \, \rm (d/1 \ Mpc)^2$,  
lasting only 1 h, where $d$ is the distance to the 
source.  Of  special interest are  the {\em gamma-ray blazars}
detected by EGRET (see e.g. \cite{Vela}).  
A detailed  study of the time structure  of 
$\gamma$-radiation for  these highly 
variable objects on timescales of several minutes
by 5@5 would provide unique information 
about the relativistic non-thermal processes 
in astrophysical jets.
An effective  operation of this instrument in the  sub-10 GeV 
regime guarantees detection of  $\gamma$-rays 
arriving from cosmological distances up to  $z \sim 3$ or so,
for which the intergalactic medium becomes almost 
transparent.  The dynamic range from several GeV 
to 100 GeV would allow important 
{\em cosmological measurements}, in particular
a study of the diffuse ultraviolet extragalactic background by 
detecting intergalactic $\gamma-\gamma$ absorption features 
in the spectra of $\gamma$-rays below 100 GeV. 
The confusion problem  (spectral cutoff due to the internal 
or extragalactic absorption ?) at such redshifts can 
probably be overcome by simultaneous observations at optical and
X-ray wavelengths. 
%
\begin{figure}[htbp]
\begin{center}
\includegraphics[width=0.5\linewidth]{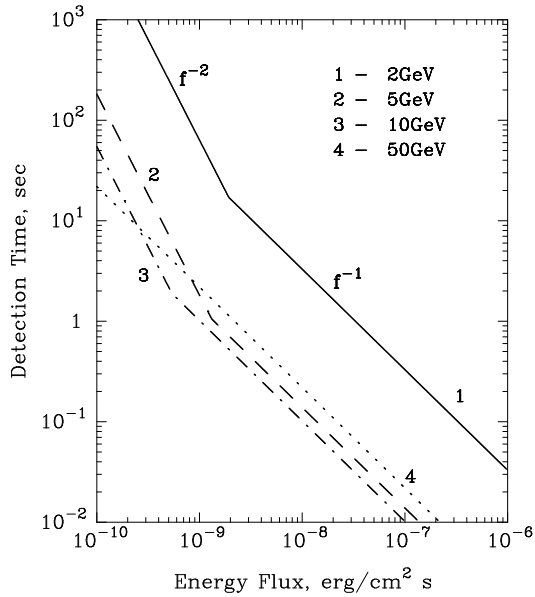}
\caption{Minimum observation time required for detection of 
$\gamma$-rays for a given energy flux (SED) $f$ 
in four energy bands centered on  $E=$2 GeV, 5 GeV, 10 GeV, and 50 GeV
with an width $\pm E/4$.}
\end{center} 
\end{figure}

5@5 can also be effectively used  
for the study of {\em galactic transient sources}, in 
particular for the detection of  short time ($\leq 1  \, \rm day$) 
$\gamma$-ray activity expected during synchrotron radio 
flares of microquasars.  

And finally,  5@5   can serve as a very powerful 
instrument for a study
of the phenomenon of {\em Gamma Ray Bursts} (GRBs). If  the 
spectra of GRBs extend to high energies without abrupt cutoffs
up to  several GeV, which is the case  at least for some 
of GRBs \cite{Hurley}, then the sensitivity of 5@5
would allow very detailed studies of the spectral and temporal 
features of GRBs in this extremely important energy region.
In Fig.~25 we show the minimum time $t_{\rm min}$ 
required for detection  of GeV $\gamma$-ray flares with a 
given energy flux $f(E)$ at 
4 different $\gamma$-ray energies: 2 GeV, 5 GeV, 10 GeV, and 50 GeV.
The calculations correspond to the 3-sigma  signal at  
each energy  $E$ within the interval  $[E-0.25 E, E+0.25 E]$,
provided that the number of detected $\gamma$-rays exceeds 10.   
In the regime of low fluxes, typically 
$f \leq 10^{-9} \, \rm erg/cm^2 s$, the $\gamma$-rays are detected 
in the presence of the heavy background induced by cosmic ray electrons. 
Therefore $t_{\rm min} \propto f^{-2}$. For fluxes larger than 
$10^{-9} \, \rm erg/cm^2 s$, the detection occurs under  almost
background-free conditions, and therefore $t_{\rm min} \propto f^{-1}$.

The results shown in Fig.~25 demonstrate the capability of 
5@5 for detection of GeV counterparts of GRBs. 
The detection of $\geq 5$ GeV episodic events  with typical  GRB 
fluxes between $10^{-8}$ and $10^{-6} \, \rm erg/cm^2 s$ would 
require only $0.1 \, \rm s$ observation time. Thus it would be possible
to monitor the spectral evolution of the source with a typical GRB 
duration from several second to 100 seconds. Remarkably, even for fluxes as low as 
$10^{-10} \, \rm erg/cm^2 s$,  the required exposure time does not exceed 
100 sec. This implies that 5@5 could serve as a unique
tool to  study GRBs in the late stages of evolution,
i.e. during the afterglows.  

5@5 is  a detector with a small field of view. 
Therefore it requires 
special strategies for the search and study of multi-GeV $\gamma$-ray 
emitters. The proximity of this energy region  to the energy range 
covered by EGRET suggests that almost all (more than 300) EGRET 
sources should also be detected by 5@5. 
The typical observation time for detection
of a ``standard'' EGRET source would  not exceed 1 hour, even if the 
spectrum of $\gamma$-rays cuts off  at energies of several GeV.   
The full overlap  of the energy range of this instrument with
the energy domain of GLAST would make  the latter a  ``the best guide''
for developing a strategy for the  study of persistent galactic and extragalactic 
objects. Generally, all sources  seen by GLAST can be potential 
targets for observations with the 5@5. 
These observations with very large 
photon statistics - not achievable by GLAST - 
could provide  detailed studies of the spectral and temporal 
features of  $\gamma$-ray sources in the  
multi-GeV region. For highly 
variable objects like blazars or galactic sources 
with relativistic jets, a  multi-wavelength approach 
including observations with radio, optical and 
X-ray detectors would be very important. 
These observations would not only inform about the pre-flaring 
or flaring states of the sources, but also would provide
complementary information for  understanding and comprehensive
modeling of the physical processes in these objects.

A special strategy should be developed for the 
search for GeV radiation from GRBs during and 
after the main event. Apparently,  
prompt information (within 10 sec or so)  
from the new generation GRB detectors like SWIFT 
(and possibly also from GLAST) would be needed,
containing  the angular coordinates 
of an  event  with an  accuracy better than $1^{\circ}$.
It would be very worthwhile   to have also a nearby 
ground-based prompt optical telescope like ROTSE 
\cite{rotse}. In  their  turn, the telescopes of 5@5 
should be rather fast in  order to be directed to the source   
not later than 1 minute after receiving the alarm from 
these  detectors.

Because of effective rejection of  hadronic showers 
by 5@5, the cosmic-ray background  
below 100 GeV is dominated by the showers 
from  cosmic-ray  electrons.
This component of electromagnetic showers 
remains  a part of the background which can be hardly   
removed, and thus 
it is the most serious limiting factor of flux sensitivities, 
especially for extended sources.
On the other hand, these electromagnetic showers with 
a known flux and spectrum of
cosmic ray  electrons, measured 
up to energies 1 TeV, can be used for 
absolute {\it energy calibration} of the instrument. 

Energy calibration with the aid of  the cosmic-ray electrons 
provides a unique tool for the {\it continuous} (on-line) control of 
the characteristics of the detector (e.g. the energy threshold, 
the detection area, {\it etc.}) during the observations. 
For example, a 10~min exposure will be enough for 
the detection of hundreds of electrons 
at any energy $E$ within $\pm E/4$ in the entire dynamical 
region from few GeV to 100 GeV (see Fig.~23).  
It is difficult to overestimate the significance of such 
calibration and control, especially for the study of
the spectral characteristics of highly  
variable $\gamma$-ray sources on sub-hour timescales.   

The basic elements of the suggested 5@5 detector
are the large optical reflectors and the multichannel
high resolution cameras. Presently, 20 m diameter
alt-azimuth mounts with the required   
precision of about 1 arcminute could be designed
and built by many companies specialized in the  construction
of large radio dishes. The area of the optical reflector
could be composed of several hundreds to thousand  
$\leq 1 \, \rm m$ diameter glass mirrors 
with protective quartz coating, quite similar to the mirrors 
used in the current or planned  imaging Cherenkov telescope  projects. 
The general requirements on  the imagers are 
a relatively large ($3^{\circ}$ or so)  field of view
with a pixel size of about $0.1^{\circ}$. This implies 
less than 1 thousand fast channels. Such a camera with
similar parameters has been already built and successfully 
operated  as part of the CAT imaging Cherenkov telescope \cite{cat}.

The decrease of  the energy threshold of the  
imaging atmospheric Cherenkov technique to several GeV
would depend to a large extent on the availability 
of exceptional sites with a dry and transparent atmosphere 
at   an altitude as high as 5 km. 
Nature does provide us with such an extraordinary site -
The Llano de Chajnantor in the Atacama desert in  Northern Chile.
This site with its very arid atmosphere was recently chosen 
for the installation of one of the most powerful future astronomical
instruments - the Atacama Large Millimeter Array (ALMA),
a funded  US-European project which will consist of 
64 12-meter radio antennas with spacing  from approximately 
150 meters to 10 km (see http://www.alma.nrao.edu/). 
The large flat area on that site could certainly accommodate
an additional  Cherenkov telescope array as well  which requires a relatively 
compact area with a radius of about  100 m. Another attractive 
feature of this site seems to be an adequate  infrastructure 
which will be built up  during the next several years for the ALMA
project.  The foreseen technological developments  of ALMA,
concerning  in particular the construction of very large
antennas operating in robotic or semi-robotic mode, could
help very much in the design of the telescopes of 5@5. 
Moreover, the neighboring Cerro Toco site \cite{tocco}
offers suitable areas at even higher altitudes, $H \simeq5.2$ and 
$5.6$ km a .s.l., with the same infrastructure advantages.  

\vspace{2mm}

\noindent
{\bf Acknowledgments}

\vspace{1mm}

We thank the anonymous referee for her/his critical
comments and remarks which helped us to improve the paper significantly. 
HQ is grateful for the award of a Presidential Chair in Science (Chile).
His research is partially funded by FONDECYT Grant No. 8970009.

\end{document}